\documentclass[showkeys,12pt, preprint,preprintnumbers,nofootinbib, groupedaddress,superscriptaddress,amsmath,amssymb]{revtex4}
\usepackage{graphicx}% Include figure files
\usepackage{dcolumn}% Align table columns on decimal point
\usepackage{bm}% bold math
\usepackage{amssymb}
\usepackage{amsmath}
\usepackage{epsfig}    
\usepackage{color}
\usepackage{slashed}
\usepackage{hhline}
\def\be{\begin{equation}}
\def\ee{\end{equation}}
\newcommand{\bea}{\begin{eqnarray}}
\newcommand{\eea}{\end{eqnarray}}
\newcommand{\nn}{\nonumber}

\def\dsl#1{#1\hspace{-2.1mm}\slash}

\numberwithin{equation}{section}

\begin{document}

{\begin{flushright}{KIAS-P16018}
\end{flushright}}

%%%%%%%%%
\title{
Confronting a  New Three-loop Seesaw Model  \\
with  the 750 GeV Diphoton Excess} 
%\preprint{KIAS-P14078}
%
\author{ P. Ko}
\email{pko@kias.re.kr}
\affiliation{School of Physics, KIAS, Seoul 02455, Korea}
\affiliation{Quantum Universe Center, KIAS, Seoul 02455, Korea}

\author{Takaaki Nomura}
\email{nomura@kias.re.kr}
\affiliation{School of Physics, KIAS, Seoul 02455, Korea}

\author{Hiroshi Okada}
\email{macokada3hiroshi@gmail.com}
\affiliation{Physics Division, National Center for Theoretical Sciences, Hsinchu, Taiwan 300}

\author{Yuta Orikasa}
\email{orikasa@kias.re.kr}
\affiliation{School of Physics, KIAS, Seoul 02455, Korea}
\affiliation{Department of Physics and Astronomy, Seoul National University, Seoul 151-742, Korea}

\date{\today}

\begin{abstract}
%We propose a new type of radiative neutrino model with a local hidden $U(1)$ symmetry, 
%in which neutrino masses are induced at the three loop level, and discuss the muon anomalous 
%magnetic moment, and dark matter candidates. Due to generalizing the hypercharge for new fields that contribute %to the neutrino masses and making them decay into  the standard model fields appropriately, a lot of nonzero %electric charges are naturally introduced. 
%Thus we show to explain the diphoton excess depending on the number of hypercharge.

We propose a new type of radiative neutrino model with a local dark $U(1)$ symmetry 
where neutrino  masses are induced at the three-loop level, and discuss the muon anomalous 
magnetic moment, and dark matter candidates therein.  By allowing the hypercharges larger 
than 3/2 for new fields that contribute  to the neutrino masses and making them decay into  
the standard model fields appropriately,  we introduce a lot of new particles with multiple 
electrical charges in a natural manner.   As a by-product, we can accommodate the 
750 GeV diphoton excess depending on the hypercharge quantum numbers of new fields 
responsible for the neutrino masses at the three-loop level.
\end{abstract}
\maketitle
\newpage

\section{Introduction}
Recently, the ATLAS and CMS collaborations reported some excess around 750 GeV 
in the observation of  the diphoton invariant mass spectrum from the run-II data at 13 TeV~\cite{ATLAS-CONF-2015-081,CMS:2015dxe}.   If confirmed, this could be a new particle 
$H$ with spin-0 or -2 and zero electric charge.
These data also indicate that 
%{\color{red} the product of the $H$ production cross section 
%and the branching fraction into the diphoton mode is about} $\rightarrow$
$\sigma ( p p \rightarrow H) \times Br (H \rightarrow \gamma \gamma) \approx 3 -10$ fb can 
explain the excess of diphoton events.
%{\color{red}  (THE CHARGED PARTICLES NEED NOT BE BOSONS)} 
Therefore, $H$ should have sizable interaction with charged particles in order to have 
a sufficiently large branching fraction in the diphoton mode.  
It implies that we have to improve the standard model (SM)  
by adding two types of particles at least:  $H$, and a new charged particle that interacts 
with $H$, and producing $\gamma\gamma$ through one loop diagram.
Along this line of thought, a large number of papers have recently been published;   
see Refs.\cite{Harigaya:2015ezk, Mambrini:2015wyu,Backovic:2015fnp,Angelescu:2015uiz,Nakai:2015ptz,Knapen:2015dap,Buttazzo:2015txu,Pilaftsis:2015ycr,Franceschini:2015kwy,DiChiara:2015vdm,Higaki:2015jag,McDermott:2015sck,Ellis:2015oso,Low:2015qep,Bellazzini:2015nxw,Gupta:2015zzs,Petersson:2015mkr,Molinaro:2015cwg,Dutta:2015wqh,Cao:2015pto,Matsuzaki:2015che,Kobakhidze:2015ldh,Martinez:2015kmn,Cox:2015ckc,Becirevic:2015fmu,No:2015bsn,Demidov:2015zqn,Chao:2015ttq,Fichet:2015vvy,Curtin:2015jcv,Bian:2015kjt,Chakrabortty:2015hff,Ahmed:2015uqt,Agrawal:2015dbf,Csaki:2015vek,Falkowski:2015swt,Aloni:2015mxa,Bai:2015nbs,Gabrielli:2015dhk,Benbrik:2015fyz,Kim:2015ron,Alves:2015jgx,Megias:2015ory,Carpenter:2015ucu,Bernon:2015abk,Chao:2015nsm,Arun:2015ubr,Han:2015cty,Chang:2015bzc,Chakraborty:2015jvs,Ding:2015rxx,Han:2015dlp,Han:2015qqj,Luo:2015yio,Chang:2015sdy,Bardhan:2015hcr,Feng:2015wil,Antipin:2015kgh,Wang:2015kuj,Cao:2015twy,Huang:2015evq,Liao:2015tow,Heckman:2015kqk,Dhuria:2015ufo,Bi:2015uqd,Kim:2015ksf,Berthier:2015vbb,Cho:2015nxy,Cline:2015msi,Bauer:2015boy,Chala:2015cev,Barducci:2015gtd,Boucenna:2015pav,Murphy:2015kag,Hernandez:2015ywg,Dey:2015bur,Pelaggi:2015knk,deBlas:2015hlv,Belyaev:2015hgo,Dev:2015isx,Huang:2015rkj,Moretti:2015pbj,Patel:2015ulo,Badziak:2015zez,Chakraborty:2015gyj,Cao:2015xjz,Altmannshofer:2015xfo,Cvetic:2015vit,Gu:2015lxj,Allanach:2015ixl,Davoudiasl:2015cuo,Craig:2015lra,Das:2015enc,Cheung:2015cug,Liu:2015yec,Zhang:2015uuo,Casas:2015blx,Hall:2015xds,Han:2015yjk,Park:2015ysf,Salvio:2015jgu,Chway:2015lzg,Li:2015jwd,Son:2015vfl,Tang:2015eko,An:2015cgp,Cao:2015apa,Wang:2015omi,Cai:2015hzc,Cao:2015scs,Kim:2015xyn,Gao:2015igz,Chao:2015nac,Bi:2015lcf,Goertz:2015nkp,Anchordoqui:2015jxc,Dev:2015vjd,Bizot:2015qqo,Ibanez:2015uok,Chiang:2015tqz,Kang:2015roj,Hamada:2015skp,Huang:2015svl,Kanemura:2015bli,Kanemura:2015vcb,Low:2015qho,Hernandez:2015hrt,Jiang:2015oms,Kaneta:2015qpf,Marzola:2015xbh,Ma:2015xmf,Dasgupta:2015pbr,Jung:2015etr,Potter:2016psi,Palti:2016kew,Nomura:2016fzs,Han:2016bus,Ko:2016lai,Ghorbani:2016jdq,Palle:2015vch,Danielsson:2016nyy,Chao:2016mtn,Csaki:2016raa,Karozas:2016hcp,Hernandez:2016rbi,Modak:2016ung,Dutta:2016jqn,Deppisch:2016scs,Ito:2016zkz,Zhang:2016pip,Berlin:2016hqw,Bhattacharya:2016lyg,D'Eramo:2016mgv, Sahin:2016lda,Fichet:2016pvq, Borah:2016uoi,Stolarski:2016dpa,Fabbrichesi:2016alj,Hati:2016thk,Ko:2016wce,Cao:2016udb,Yu:2016lof,Ding:2016ldt,Alexander:2016uli,Davis:2016hlw,Dorsner:2016ypw,Faraggi:2016xnm,Djouadi:2016eyy,Ghoshal:2016jyj,Nomura:2016seu,Chao:2016aer,Han:2016bvl,Okada:2016rav,Franzosi:2016wtl,Martini:2016ahj,Cao:2016cok,Chiang:2016ydx,Aydemir:2016qqj,Abel:2016pyc,Harland-Lang:2016qjy,King:2016wep,Kawamura:2016idj,Kavanagh:2016pso,Nomura:2016rjf,Geng:2016xin,Bertuzzo:2016fmv,Ben-Dayan:2016gxw,Barrie:2016ntq,Aparicio:2016iwr,Ding:2016udc,Li:2016xcj,Salvio:2016hnf,Ge:2016xcq,Giddings:2016sfr,Ellwanger:2016qax,Arbelaez:2016mhg,Bae:2016xni,Gross:2016ioi,Han:2016fli,Hamada:2016vwk,Dasgupta:2016wxw,Goertz:2016iwa,Zhang:2016xei,Staub:2016dxq,Baek:2016uqf,Cvetic:2016omj}.

Another motivation for going beyond the SM (BSM) comes from nonzero 
neutrino masses and mixings as well as nonbaryonic cold dark martter, for which  there are 
a huge number of different models.  For neutrino masses and mixings,  radiative seesaw 
models are renowned as having an elegant mechanism to explain  tiny neutrino masses 
within renormalizable theories.  Some kinds of radiative neutrino mass  models have new 
charged particles that are naturally introduced as a mediating particles 
in the loops responsible for neutrino masses and mixings. Moreover some of the radiative 
neutrino mass models  can accommodate  dark matter (DM) candidates, which would  
clearly be an advantage, since one can explain both neutrino masses and mixings and 
nonbaryonic DM in one framework.
% that also plays 
%an important role in generating tiny neutrino masses.} 

In this paper, we propose a new radiative seesaw model with a local
 dark $U(1)$ symmetry, where neutrino masses and mixings are 
generated at the three-loop  level, and DM candidates are introduced naturally in the model.  
Then we also explain  the muon anomalous magnetic moment, the relic density of our two 
DM candidates (Majorana fermion  and/or scalar), 
as well as the recent 750 GeV diphoton excess. Notice here that any lepton flavor violating 
processes can easily be evaded by diagonalizing the Yukawa term that induces the muon 
anomalous magnetic moment, since our neutrino masses have another Yukawa coupling 
($g_L, g_R$)  (see Eq.~(\ref{Eq:lag-flavor}) below). 
Therefore, the neutrino mixing is expected to be generated via $g_L$ and $g_R$.  
%%%
Since both of the DM candidates have the local dark $U(1)_X$ charge,  they interact with the dark
neutral vector boson $Z'$, which plays an important role in the DM thermal relic density 
in this paper. And we can easily evade the constraint for a DM direct detection search 
such as LUX~\cite{Akerib:2013tjd},  assuming that the kinetic mixing between $Z^{'}$ 
and the SM U(1)$_Y$ gauge  field is small enough.
Moreover, since our model generalizes the hypercharge of isospin doublet fields 
as well as isospin singlet fields without violating the structure of neutrinos, 
a lot of nonzero electric charged fields can be involved in our theory. 
Thus, we can explain the diphoton excess naturally, depending on the hypercharge quantum numbers of new particles. 
However, in general, allowing such a general range in hypercharge number could cause 
a problem of stable charged particles. 
Therefore, we have to make them decay into the SM (or DM) appropriately. 
In order to realize this, we add some more nonzero charged bosons and show 
the appropriate  decay processes for each value of hypercharge, retaining our model. 
Then such new bosons shall also play a role in contributing the diphoton excess.

This paper is organized as follows. 
In Sec.~II, we define our model for three-loop neutrino masses and DM, the mass matrices for 
the neutral scalar bosons and neutral fermions including the DM candidates, and the decay 
properties of exotic particles. 
In Sec.~III, we discuss the lepton flavor physics, focusing on  the radiative generation of 
neutrino masses at three-loops, the muon $(g-2)_\mu$ within our model, and charged lepton 
flavor violation.   In our model there are two candidates for cold DM, one bosonic and the 
other fermionic. In Sec.~IV, the phenomenology of these two DM candidates is discussed. 
In Sec.~V, we discuss the 750 GeV  diphoton excess within this model in detail. 
Finally we summarize the results in Sec.~VI.
%In appendices, we show the explicit Higgs potential and ...
%\newpage

%%%%%%%%%%%%%%%%%%%%%%%%%%%%%%%%%%%%%
 \section{Model and Particle properies}
 \subsection{Particle contents and the model Lagrangian}
 \begin{widetext}
\begin{center} 
\begin{table}[tb]
%\begin{tiny}
\begin{tabular}{|c||c|c|c|c||c|c|c|c|c|}\hline\hline  
&\multicolumn{4}{c||}{Lepton Fields} & \multicolumn{5}{c|}{Scalar Fields} \\\hline
& ~$L_L$~ & ~$e_R^{}$~ & ~$L'^{}_{}$  ~ & ~$N_{}$~  & ~$\Phi_{}$~ & ~$\Phi'$  & ~$S^{+q}$  & ~$S^{}$   & ~$\varphi$ \\\hline 
$SU(2)_L$ & $\bm{2}$  & $\bm{1}$  & $\bm{2}$ & $\bm{1}$ & $\bm{2}$ & $\bm{2}$  & $\bm{1}$   & $\bm{1}$   & $\bm{1}$ \\\hline 
$U(1)_Y$ & $-\frac12$ & $-1$  & $-\frac{N}2$ & $0$ & $\frac12$ & $\frac{N}2$ & $q$  & $0$   & $0$  \\\hline
$U(1)_X$ & $0$ & $0$  & $x$  & ${x}$ & $0$ & ${0}$ & $-x$  & $x$   & $-2x$  \\\hline
% $Z_2$ & $+$ & $+$  & $+$  & ${-}$ & $+$ & ${+}$ & $+$  & $-$ & $+$  & $+$  \\\hline
\end{tabular}
\caption{Contents of fermion and scalar fields
and their charge assignments under $SU(2)_L\times U(1)_Y\times U(1)_X$, where $q\equiv \frac{N-1}{2}$ and $(3\le)N$ is an arbitrary odd number. 
Note that we have introduced three generations of new fermions $L'$ and $N$, whereas  
only one set of the listed scalar contents are introduced in the scalar sector. }
\label{tab:1}
% \end{tiny}
\end{table}
\end{center}
\end{widetext}
%
%We discuss a two-loop induced radiative neutrino model. 
In this section, we explain our model for three-loop neutrino masses with new particles 
that are charged under a dark $U(1)_X$ symmetry 
 as well as the SM $SU(2)_L \times U(1)_Y$ gauge symmetry. 
The particle contents and their gauge charges are shown in Table~\ref{tab:1}.
 Let us note that all the new particles are color-singlets. 
To the SM, we have added  vector-like exotic isospin doublet fermions $L'$ 
with a weak hypercharge equal to $Y = -N/2$,  %hypercharge, 
SM singlet Dirac fermions $N$, an isospin doublet boson $\Phi'$ with $Y = N/2$, 
an isospin singlet scalar $S^{\pm q}$ with electric charge $Q=q$, and two isospin 
singlet neutral scalars $S^{}$ and  $\varphi$ that carry different $U(1)_X$ charges.
We assume that $U(1)_X$ is spontaneously broken by the nonzero vacuum expectation value (VEV) of  
a $U(1)_X$-charged SM singlet scalar $\varphi (x)$. 
Notice here that $N (\ge 3)$ is an arbitrary odd number \footnote{ 
%WHY IS N ODD ? LET US MAKE A NOTE HERE ABOUT WHAT HAPPENS FOR EVEN N..
For even N, the electric charges of components in $L'$ and $\Phi'$ become half-integer, where the lightest particle with half-integer charge cannot decay. } 
and $q\equiv \frac{N-1}{2}$ 
is an integer.   Thus, the electric charges of each component of $L'$ and $\Phi'$ are 
$(-q,-q-1)$ and $(q+1,q)$, respectively.  Therefore, we shall define  
$L'\equiv ( E^{-q},E^{-q-1} )^T$ and $\Phi' \equiv ( \phi^{1+q},\phi^{+q} )^T$ in the following. 

Then the  renormalizable parts of the relevant Yukawa interaction Lagrangian and the scalar 
potential  under these gauge symmetries are given by 
\begin{align}
-\mathcal{L}_{Y}
&=
y_{\ell_{ij}} \bar L_{L_i} \Phi e_{R_j} + f_{ij} \bar L_{L_i} L'_{R_j} S^{+q}
+g_{L_{ij}} \bar L'_{L_i}\tilde \Phi N_{R_j}   +g_{R_{ij}} \bar L'_{R_i}\tilde \Phi N_{L_j} \nn\\
&  + \frac{y_{N_{L_{i}}} }{2} \varphi \bar N^c_{L_i} N_{L_i}    + \frac{y_{N_{R_{i}}} }{2} \varphi \bar N^c_{R_i} N_{R_i} + M_{D_{ij}} \bar N_{L_i} N_{R_j}
+ { M_{L_i}} \bar L'_{L_i} L'_{R_i}  +{\rm c.c.},\label{Eq:lag-flavor} 
\\
V =& \ m_S^2 |S|^2 + m_\varphi^2 |\varphi|^2 + m_{S^\pm}^2 |S^{+q}|^2 + m_{\Phi}^2 |\Phi|^2 + m_{\Phi'}^2 |\Phi'|^2
  +  \frac{\mu}2 (\varphi S^2 +{\rm c.c.})  \nonumber \\
&
+ \kappa \left( (\Phi'^\dag \Phi) S^{+q} S + {\rm c.c.} \right)
 + \lambda_S |S|^4 + \lambda_\varphi |\varphi|^4 + \lambda_{S^\pm} |S^{+q}|^4 + \lambda_{\Phi} |\Phi|^4 + \lambda_{\Phi'} |\Phi'|^4
   \nonumber \\
& 
+ \lambda_{S\varphi} |S|^2 |\varphi|^2 + \lambda_{S S^\pm} |S|^2 |S^{+q}|^2 + \lambda_{S \Phi} |S|^2 |\Phi|^2  
+ \lambda_{S \Phi'} |S|^2 |\Phi'|^2  + \lambda_{\varphi S^\pm} |\varphi|^2  |S^{+q}|^2  \nonumber \\
& + \lambda_{\varphi \Phi} |\varphi|^2  |\Phi|^2   + \lambda_{\varphi \Phi'} |\varphi|^2  |\Phi'|^2  
  + \lambda_{S^\pm \Phi} |S^{+q}|^2  |\Phi|^2   + \lambda_{S^\pm \Phi'} |S^{+q}|^2  |\Phi'|^2   + \lambda_{\Phi\Phi'} |\Phi|^2  |\Phi'|^2 ,  
\label{eq:potential} 
\end{align}
where we take $y_{N_{L/R}}$ in the diagonal basis without loss of generality. 

We assume that only the SM Higgs doublet $\Phi_{}$  and the $U(1)_X$-charged SM 
singlet scalar$\varphi$ have nonzero VEVs, 
which are denoted by $v/\sqrt2$ and $v'/\sqrt2$ respectively. 
And we obtain the Majorana masses  $M_{N_{L/R}}\equiv y_{N_{L/R}} v'/\sqrt2$.
The first term of $\mathcal{L}_{Y}$ generates the SM
charged-lepton masses $m_\ell\equiv y_\ell v_1/\sqrt2$ after the spontaneous breaking of electroweak symmetry by $\langle \Phi \rangle = v/\sqrt{2}$.
We work in the basis where all the coefficients are real and positive for simplicity. 
In the unitary gauges, one has 
\[
\Phi^T =( 0 ,\frac{v+h}{\sqrt2} ) ,  \ \ \ 
\varphi=\frac{v'+h'}{\sqrt2},
\]
where the CP-odd component of $\varphi$ is absorbed by the longitudinal component $Z'$ as NG boson.

The nonzero $U(1)_X$ quantum number $x\neq0$ is arbitrary, but its assignment for each 
field is unique so that we can realize our three-loop neutrino model. And there exists a remnant 
$Z_2$ symmetry ($S \rightarrow -S$ in Eq. (II.1)) from the $\mu$-term  even after the 
spontaneous breaking of dark $U(1)_X$ symmetry via $\langle \varphi \rangle$, 
which plays a role in assuring the stability of the dark matter candidate~\cite{Baek:2014kna}. 
Therefore, the dark matter candidate in our model is the lightest mass eigenstate of the 
Dirac neutral  fermion $N_{L/R}|_{\rm lightest} = X$ and/or the lightest isospin singlet boson 
of $S\equiv (S_R+i S_I)/\sqrt2$.   Here we identify the first generation of the mass eigenstate of 
$N_{L/R}$ or $S_I$ as a dark matter candidate.   
In addition, we have a massive $Z'$ boson which is associated with $U(1)_X$ after the symmetry breaking.
%{\color{blue}
%}

%After spontaneous breaking of the local $U(1)_X$ by $\langle \varphi \rangle =v'/\sqrt{2}$, %
%{\color{red} 
%The isospin doublet scalar fields can be parameterized as $\Phi=[w^+,\frac{v+h+iz}{\sqrt2}]^T$
%where $v~\simeq 246$ GeV is VEV of the Higgs doublet, and $w^\pm$
%and $z$ are respectively absorbed by the longitudinal component of $W$ and $Z$ boson.
%The isospin singlet scalar field can be parameterized by $\varphi=\frac{v'+h'+i z'}{\sqrt2}$ where %$z'$ is absorbed by the longitudinal component of $Z'$ boson. 
%(DO WE NEED PARTS ??? THIS IS A BASIC, AND WE CAN REMOVE IT,  
%AND REPLACE IT BY THIS NEW EQUATIONS.)}

\subsection{Mass matrices for neutral scalar bosons and neutral fermions}
The mass matrix for the CP-even neutral scalar Higgs bosons is given by 
\begin{equation}
\frac{1}{2} \begin{pmatrix} h' & h \end{pmatrix} M^2 \begin{pmatrix} h' \\ h \end{pmatrix} 
=  \frac{1}{2}\begin{pmatrix} h' & h \end{pmatrix} \begin{pmatrix} \tilde m_{h'}^2 & \lambda_{\varphi \Phi} v v' \\ \lambda_{\varphi \Phi} v v' & \tilde m_h^2 \end{pmatrix} \begin{pmatrix} h' \\ h \end{pmatrix}
\end{equation}
where $\tilde m_h = \sqrt{2 \lambda_\Phi} v$ and $\tilde m_{h'} = \sqrt{2 \lambda_\varphi} v'$.
Then, the mass eigenstates are defined by 
\begin{equation}
\begin{pmatrix} h' \\ h \end{pmatrix} = \begin{pmatrix} \cos \alpha & - \sin \alpha \\ \sin \alpha & \cos \alpha \end{pmatrix} \begin{pmatrix} H \\ h_{\rm SM} \end{pmatrix},
\end{equation}
where the scalar mixing angle $\alpha$ satisfies the following relation:
\begin{equation}
\label{eq:alpha}
\tan 2 \alpha = \frac{2 \lambda_{\varphi \Phi} v v'}{(\tilde m_{h'}^2 - \tilde m_h^2)}.
\end{equation}
Here $h_{\rm SM}$ and $H$ denote the SM Higgs and the heavier new CP-even Higgs, respectively.
\footnote{In Sec. V, the scalar $H$ will be identified as the scalar boson that is  responsible for the 750 GeV diphoton execss. }
Then the mass eigenvalues are 
\begin{equation}
 m^2_{h_{\rm SM}, H} = \frac{1}{2} \left(\tilde m^2_{h'} + \tilde m^2_{h}  \mp \sqrt{(\tilde m^2_{h'} - \tilde m^2_{h})^2 + 4 \lambda_{\lambda_{\varphi \Phi}}^2 v^2 v'^2} \right)\,.
\end{equation}
The mass of $Z'$ is also given by
\begin{equation}
\label{eq:Z'mass}
m_{Z'} = 2 x g_X v',
\end{equation}
where we have ignored the $Z$-$Z'$ mixing effect, assuming kinetic mixing is negligibly small.
%In our analysis of $H$ production via gluon fusion, 
The gluon fusion process of $H$ production is induced by mixing with SM Higgs where  
we focus on the Yukawa interactions of $H$ and the top quark  as
%%%
\begin{align}
{\cal L}^Y \supset &  -   \frac{m_{t} \sin \alpha}{v} \bar t t H.
\end{align}

%{\it Neutral Fermion mass matrix}:
%%%
The isospin singlet exotic neutral fermion mass matrix is given by 
\begin{align}
-{\cal L}_{\text{mass}} &= 
(\overline{N^{c}_L},\overline{N^{}_R})
\begin{pmatrix}
M_{N_L} &   M_D \\
M^\dag_D & M_{N_R}
\end{pmatrix}
\begin{pmatrix}
N^{}_L \\
N^{c}_R
\end{pmatrix} + \text{h.c.}=
(\overline{N_1^{c} },\overline{N_2^{}})
\begin{pmatrix}
M_{N^1} &   0 \\
0  & M_{N^2}
\end{pmatrix}
\begin{pmatrix}
N_{1}^{} \\
N_{2}^{c}
\end{pmatrix} + \text{h.c.} , 
\end{align}
where we define $M_{N_{L/R}}\equiv y_{N_{L/R}} v'/\sqrt2$. % and assume $h^T=h$. 
%%%
In general, the diagonalization is very complicated because $M_D$ is the general 3$\times 3$ matrix. 
However, once we take $M_D$ as the diagonal basis similar to the $M_{N_{L/R}}$ terms, 
we can simplify this sector and consider one flavor basis. Hereafter, we adapt this assumption for simplicity. 
%%%
Then the mass eigenstates $N_1$ and $N_2$ are defined by the following transformation:
\begin{align}
\begin{pmatrix}
N^{}_L \\
N^{c}_R
\end{pmatrix}
=
\begin{pmatrix}
c_{\theta_N} & -s_{\theta_N} \\
s_{\theta_N} &c_{\theta_N}
\end{pmatrix}
\begin{pmatrix}
N_1^{} \\
N_2^{c}
\end{pmatrix}, 
\end{align}
where we define $s_{\theta_N} \equiv \sin\theta_N$ and $c_{\theta_N} \equiv \cos\theta_N$.
The mass eigenvalues ($M_{N^{1}}< M_{N^{2}}$) and the mixing angle $\theta_N$ are, respectively, given by 
\begin{align}
M_{N^{1,2}} &= \frac{1}{2}\left(M_{N_L} + M_{N_R} \mp\sqrt{(M_{N_L} - M_{N_R})^2 + 4M_D^2}\right), \quad
\tan2  \theta_N  = \frac{2 M_D}{M_{N_L} - M_{N_R}}. 
\end{align}
Furthermore, we define $\Psi_1\equiv N_1+N_1^c\ (\bar\Psi_1\equiv \overline{N_1}+\overline{N_1^c})$ and 
$\Psi_2\equiv N_2^c+N_2\ (\bar\Psi_2\equiv \overline{N_2^c}+\overline{N_2})$ for convenience.
So we rewrite  our Lagrangian in terms of $\Psi_i(i=1,2)$, where  the transformation rules are given by 
\begin{align}
N_1= P_L\Psi_1, \ N_1^c= P_R \Psi_1, \ N_2= P_R \Psi_2, \ N_2^c= P_L\Psi_1,
\end{align}
where the mass eigenstate is the same as that of $N_{1/2}$.

\subsection{Decay properties of exotic particles}
%{\it Decay processes}:\\
Now we consider the decay processes for the newly introduced exotic particles.
Regardless of the electric charge $q$, the particle $E^{-q-1}$ always decays into $E^{-q}$ 
and the charged gauged boson $W^-$.    
And $E^{-q}$ decays into $S^{-q}$ and active neutrinos if $E^{-q}$ is heavier than $S^{-q}$,
or  $S^{-q}$ decays into $E^{-q}$ and active neutrinos if $E^{-q}$ is lighter than $S^{-q}$.
%%%
Moreover, $\phi^{-q}$ can decay into $S^{-q} + S$ or $E^{-q} + \Psi_i$ (with the missing $E_T$ 
generated by $S$ or $\Psi_i$), depending on the mass hierarchies among the particles involved. 
In order to simplify the analysis, we just assume  
$m_{S^{\pm q}}\ {\rm +\ missing}\ {\rm or}\ M_L  \ {\rm +\ missing}< m_{\phi^{\pm q}}$.
Therefore, all we have to find to take care of the decay is how to make the $S^{\pm q}$ or $E^{\pm q}$ 
decay into the SM particles, which depends on the quantum number $N$. Thus we classify the 
model in terms of the concrete number of $N$ below.
We also symbolize additional fields to contribute to the decay as $D^{\pm q}$.
Notice here that $N$ starts from 3, since we assume $q\neq0$.

\subsubsection{N=3}
This is equivalent to $q=1$. 
In this case, the model is identified as the previous work in Ref.~\cite{Okada:2015hia},
and additional  new fields are not needed. But since sizable muons $(g-2)_\mu$ cannot be 
obtained within the 3.2$\sigma$ level as shown in Sec.~\ref{sec:g-2}, we do not consider this case further.

\subsubsection{N=5}
This is equivalent to $q=2$. 
In this case, by introducing a new field $D^{\pm}$ that is an isospin singlet and singly charged 
boson with $\mp x$ $U(1)_{X}$ charge,  we can add the terms  
\begin{align}
-{\cal L}_{\rm new} \approx   g' \bar N^c_R e_R  D^{+}  + y_{eE}  S^{++} D^{-} D^{-} S^* 
+ {\rm c.c.},
\end{align}
and then the decay processes are as follows:
\begin{align}
S^{--}\to   2 D^{-} (+ S^*) \to 2\ell^- + 2 N (+ S^*),
 \end{align}
where $S$ and $N$ are expected to appear as missing energy signatures at colliders.

\subsubsection{N=7}
This is equivalent to $q=3$. 
In this case, by introducing two new isospin singlet fields $D^{\pm}$  with $\pm x$ $U(1)_{X}$ charge 
and  $D^{\pm\pm}$  with neutral $U(1)_{X}$ charge,
we can add the terms  
\begin{align}
-{\cal L}_{\rm new} \approx    g' \bar N_L e_R  D^{+}  +  g'' \bar e^c_R e_R  D^{++}  + \lambda'  S^{+++} D^{--} D^{--} D^+ + {\rm c.c.},
\end{align}
%where $S^{\pm\pm}$ plays as a role in generating the decaying processes for the exotic fields only.
and then the decay processes are as follows:
\begin{align}
S^{---}\to   2 D^{--} + D^+  \to 4\ell^- + \ell^+ +N.
 \end{align}

\subsubsection{N=9}
This is equivalent to $q=4$. 
In this case, by introducing a new isospin singlet boson $D^{\pm\pm}$  with neutral $U(1)_{X}$ charge,
we can add the terms  
\begin{align}
-{\cal L}_{\rm new} \approx   g'' \bar e^c_R e_R  D^{++}  + \lambda'  S^{++++} D^{--} D^{--} S + {\rm c.c.},
\end{align}
where additional fields play a role in generating the decaying processes for the exotic fields only.
Then the decay processes are as follows:
\begin{align}
S^{----}\to   2 D^{--} (+ S) \to 4\ell^-(+ S).
 \end{align}

\subsubsection{N=11}
This is equivalent to $q=5$. 
In this case, by introducing two new isospin singlet fields $D^{\pm}$  with $\mp x$ $U(1)_{X}$ charge 
and  $D^{\pm\pm}$  with neutral $U(1)_{X}$ charge,
we can add the terms  
\begin{align}
-{\cal L}_{\rm new} \approx    g' \bar N^c_R e_R  D^{+}  +  g'' \bar e^c_R e_R  D^{++}  + \lambda'  S^{+++++} D^{--} D^{--} D^- + {\rm c.c.},
\end{align}
%where $S^{\pm\pm}$ plays as a role in generating the decaying processes for the exotic fields only.
and then the decay processes are as follows:
\begin{align}
S^{-----}\to   2 D^{--} + D^-  \to 5\ell^- +N.
 \end{align}

\begin{figure}[tb]
\begin{center}
\includegraphics[width=13cm]{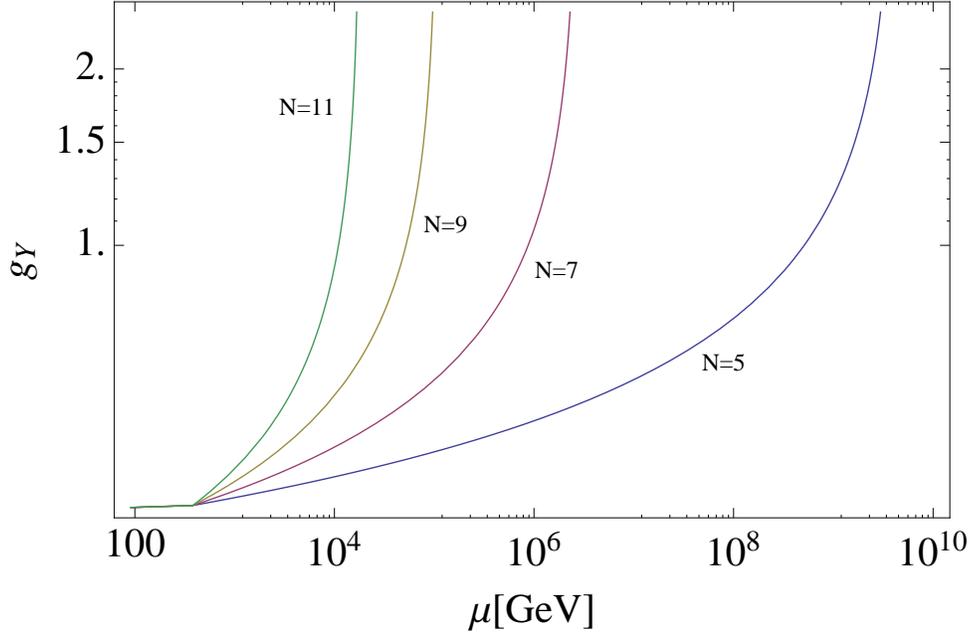}
\caption{The running of $g_Y$ in terms of a reference energy of $\mu$, depending on each of $N=5,7,9,11$.}
\label{fig:rge}
\end{center}\end{figure}

It is worthwhile to mention the Landau pole for $g_Y$ in the presence of new exotic fields 
with nonzero hypercharge.~\footnote{ The potential problem of a Landau pole at low energies associated to the diphoton excess is also discussed in e.g. Refs.~\cite{Gross:2016ioi,Goertz:2016iwa,Staub:2016dxq}.}   The new beta function of $g_Y$ for $SU(2)_L$ doublet fields 
with $\pm N/2$ hypercharge is given by
\begin{align}
\Delta b^f_Y=N^2 \ ,\quad \Delta b^b_Y=\frac{N^2}{6} \ ,
\end{align}
where the upper indices of $\Delta b$ represent the fermion (f) and the boson (b), respectively.
Similarly, the beta function for the $SU(2)_L$ singlet boson with $(N-1)/2$ hypercharge 
is given by 
\begin{align}
\Delta b^s_Y=\frac{(N-1)^2}{12} \ .
\end{align}
We include contributions from exotic doublet fermions $L'$, a new doublet scalar $\Phi'$, 
a charged singlet scalar $S^{\pm q}$, and additional singlet charged scalars for each $N$.
The resultant flow of $g_Y$ is then given by Fig.~\ref{fig:rge} for each velue of $N$, where 
$\mu$ is a reference energy.
Moreover, we fix the threshold to be the mass of the SM $Z$ boson, and we assume that the masses of all the fields 
contributing to the beta function are 380 GeV. 
This suggests that our model is valid 
up to  the scale of ${\cal O}$ (10 TeV) even if we take $N=11$. 

\section{Neutrino masses at three-loop level and the muon $(g-2)_\mu$}

%\subsection{Neutrino mass matrix}
%%%%%%%%%%%%%%%%%%%
%\if0
\begin{figure}[tb]
\begin{center}
\includegraphics[width=80mm]{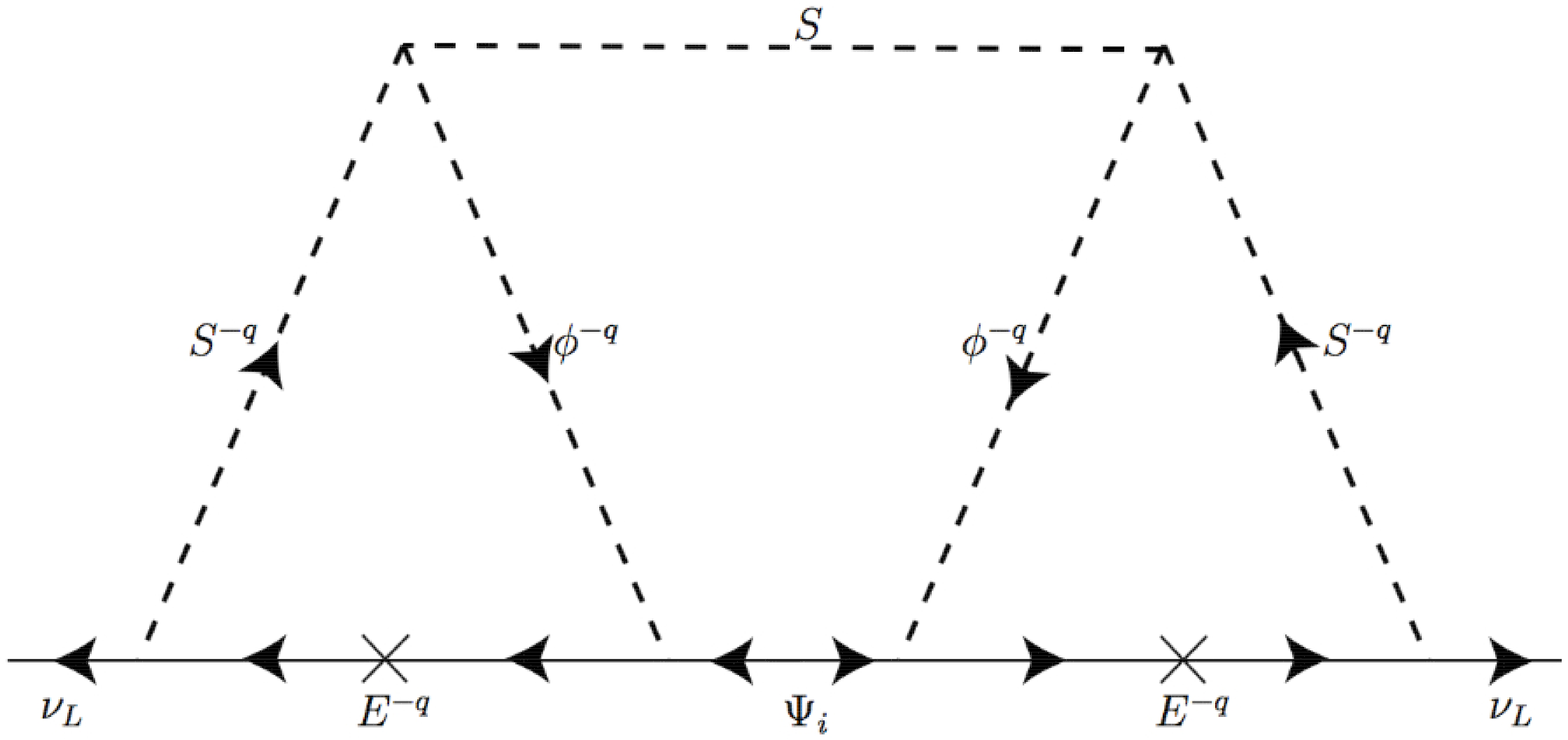}
\includegraphics[width=80mm]{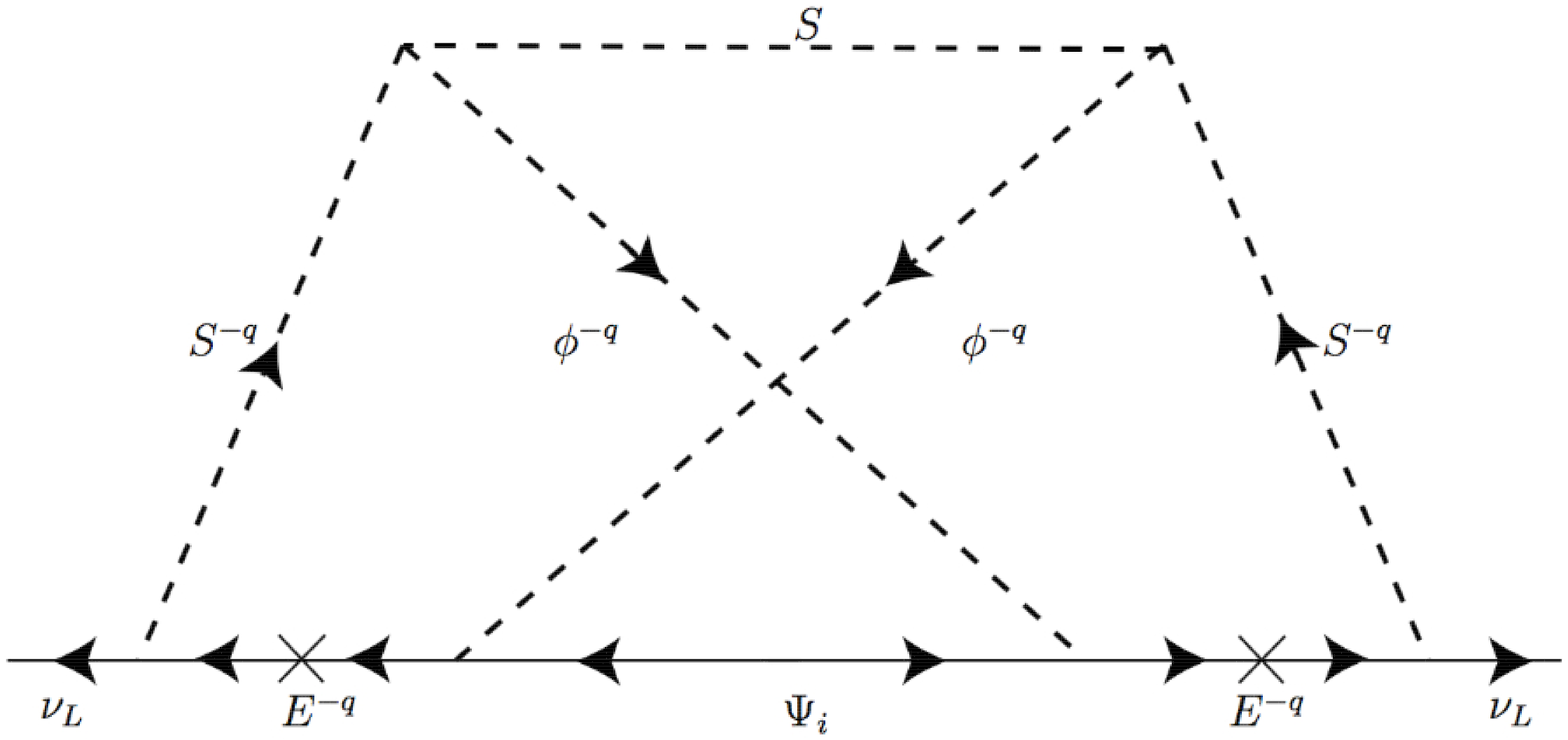}
\includegraphics[width=80mm]{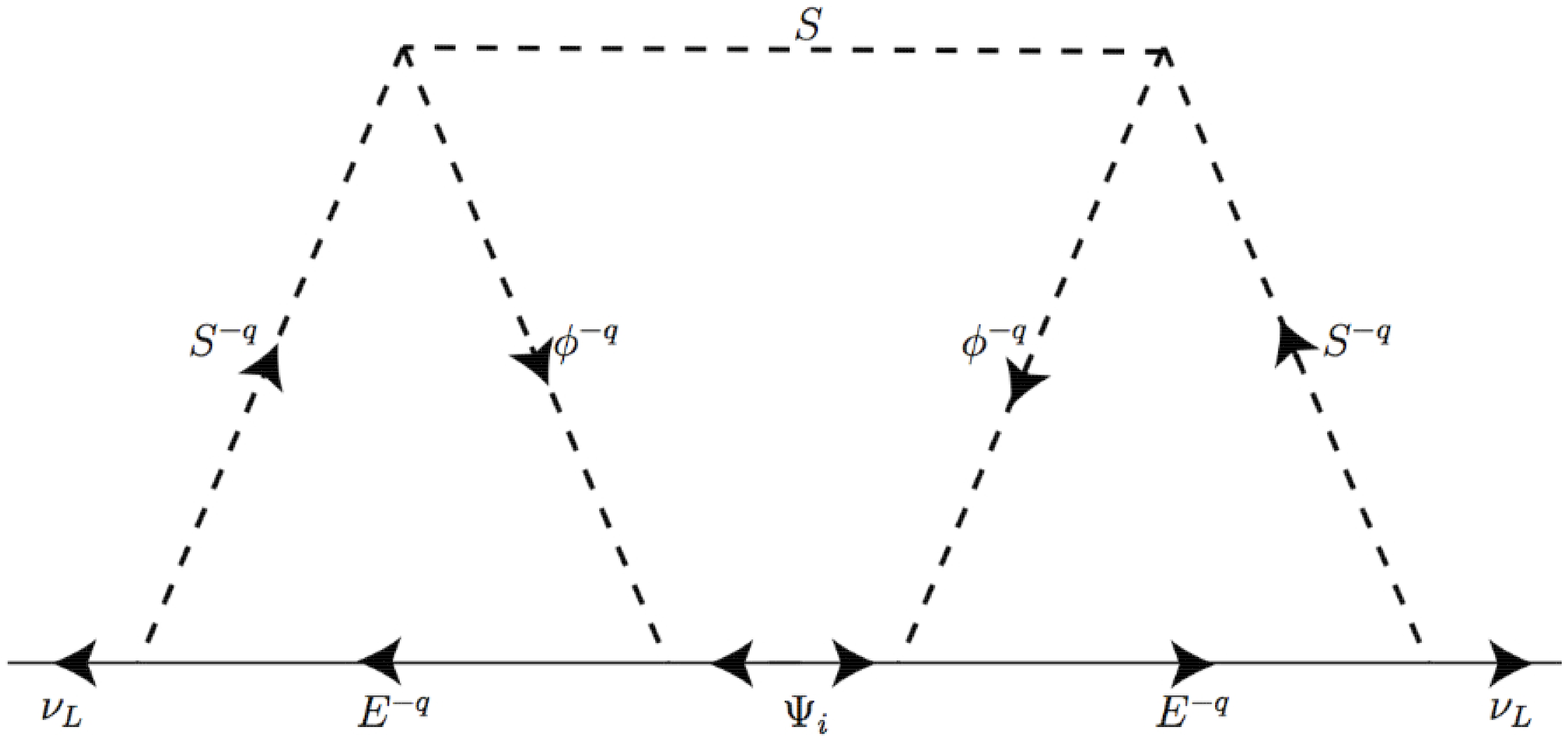}
\includegraphics[width=80mm]{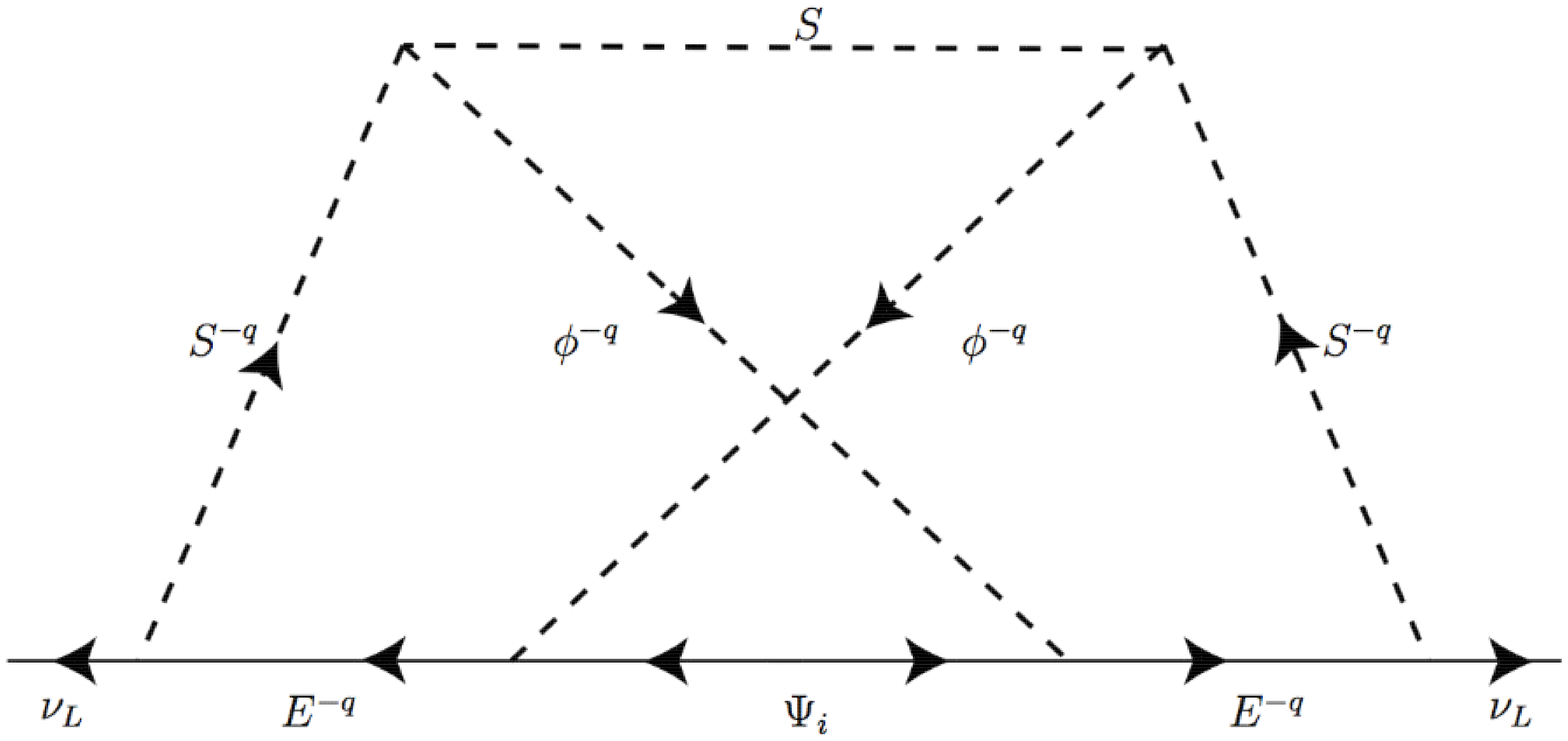}
\caption{ Neutrino mass matrix at the three-loop level, where the top-left figure corresponds to $m^{I}_\nu$, 
the top-right figure corresponds to $m^{II}_\nu$, the bottom-left figure corresponds to $m^{III}_\nu$, and
the bottom-right figure corresponds to $m^{IV}_\nu$. 
 The arrows in the diagrams indicate chirality flow for neutral fermion lines, 
electric charge flow for boson lines, and both flows for charged fermion lines.
%%%
}   \label{fig:neut1}
\end{center}\end{figure}
%%%%%%%%%%%%%%%%%%%

%\fi
\subsection{Neutrino mass matrix at three-loop level}
%{\it Neutrino mass matrix}:

%%%
Within the model Lagrangian described in the previous section, we are now ready to discuss 
the neutrino masses at the three-loop level.
The leading contribution to the active neutrino masses $m_\nu$  in our model arises at 
the three-loop level as shown in Fig.~\ref{fig:neut1}, and its formula is given as follows:
%\begin{widetext}
\begin{align}
(m_{\nu})_{ij}& \equiv (m_{\nu}^I)_{ij} +  (m_{\nu}^{II})_{ij} +   (m_{\nu}^{III})_{ij} +   (m_{\nu}^{IV})_{ij} ,\\
%%%
(m_{\nu}^{I})_{ij} &=
 \frac{\kappa^2 v^2  }{2(4\pi)^6 M^4_{\rm max} } 
\sum_{\alpha,\beta,\gamma=1}^3 (f_{i\alpha} M_{L_\alpha} g_{L_{\alpha\beta}}  g_{L_{\gamma\beta}} M_{L_\gamma} f_{j\gamma})
\left[ c^2_{\theta_N} M_{\Psi_{1\beta}} G_{I} (X_{\Psi_{1\beta}}) +  s^2_{\theta_N} M_{\Psi_{2\beta}} G_{I} (X_{\Psi_{2\beta}})\right] ,\nn\\
%%%
(m_{\nu}^{II})_{ij} &=
 \frac{\kappa^2 v^2  }{2(4\pi)^6 M^2_{\rm max} } 
\sum_{\alpha,\beta,\gamma=1}^3 (f_{i\alpha}  g_{R_{\alpha\beta}}  g_{R_{\gamma\beta}} f_{j\gamma})
\left[ s^2_{\theta_N} M_{\Psi_{1\beta}} G_{II} (X_{\Psi_{1\beta}}) +  c^2_{\theta_N} M_{\Psi_{2\beta}} G_{II} (X_{\Psi_{2\beta}})\right] ,\nn\\
%%%
(m_{\nu}^{III})_{ij} &=
 (m_{\nu}^{I})_{ij} (G_I \to G_{III}) , \nn\\
(m_{\nu}^{IV})_{ij} & =
 (m_{\nu}^{II})_{ij} (G_{II} \to G_{IV}) ,
\end{align}
where we have defined  $X_f\equiv (m_f/M_{\rm max})^2$, and $M_{\rm max}={\rm Max}[M_L,M_{\Psi_i}, m_{S^\pm},m_{S^{\pm5}},m_{R},m_{I}]$. 
The loop functions $G_{I-IV}$ are given in the Appendix.
%%%
The neutrino masses $m_\nu$ should be 
\[
0.001\ {\rm eV}\lesssim m_\nu \lesssim 0.1\ {\rm eV}
\]
from the neutrino oscillation data~\cite{pdf}.

Let us discuss what is new and unique in our model for generating the active neutrino mass 
matrix at three-loop level, compared with other three-loop models in the literature~\cite{Krauss:2002px,Aoki:2008av, Gustafsson:2012vj}. 
%{\color{red} PLEASE ADD HERE SOME REFERENCES FOR 3-LOOP NEUTRINO MASSES PROPOSED IN EARLIER LITERATURE}. 
A new part of this model introduces a set of isospin doublet fermions $L^{'}$ 
and an isodoublet  scalar boson $\Phi^{'}$, both of which have large hypercharges 
$Y = \pm N/2$ (with  $3\le N)$ (see Table I) in order to induce the active neutrino masses 
at the three-loop level. 
In this case, however, it would generally be difficult to make them decay into the SM fields appropriately due to specific charges. 
%{\color{red} , since there are few kind of models.  (CAN WE REMOVE THIS RED PART ???)}
To solve this problem, we also introduce a set of isospin singlet fermions $N$'s and a isospin  singlet scalar boson $S$, both of which can be a DM candidate. 

All these new isosinglet fields can also play a role in generating the neutrino masses by 
connecting the isospin doublet exotic fields. 
%%%  
Its connection is realized by the local dark $U(1)_X$ symmetry, which is 
one of the remarkable and interesting features of our model.
The model presented in this paper is the first proposal for a three-loop seesaw neutrino 
model with a dark sector and local dark gauge symmetry. 
%There exist no such kind of three loop induced neutrino models with local hidden $U(1)$ %symmetry.   with $\pm N/2$ hypercharge.
Thus, one can obtain a sizable neutrino mass scale by controlling these exotic masses.
%~\footnote{\color{red}  Any three loop models that are proportional to charged lepton masses of SM~\cite{Aoki:2008av} could  actually be few parameter spaces to satisfy the current neutrino oscillation data nowadays. (WHAT IS THE MEANING OF THIS FOOTNOTE ???)} 
Moreover, since one can generalize the hypercharges of isospin doublet fields,   their 
electric charges can be increased arbitrarily. 
Thus, we can explain the muon anomalous magnetic moment, as well as the 750 GeV diphoton excess 
from the loops involving new particles with large electric charges, as we will discuss later. 
The local dark symmetry also plays an important role in explaining the measured relic 
density of DM. In this sense, we emphasize that all the phenomenology such as the muon 
anomalous magnetic moment, the DM property, and the 750 GeV diphoton excess, are strongly 
correlated to the neutrino masses, which are quite new features to discriminate this approach 
from other radiative models.

\subsection{Muon anomalous magnetic moment and charged lepton flavor violation \label{sec:g-2}}

\begin{figure}[tb]
\begin{center}
\includegraphics[width=60mm]{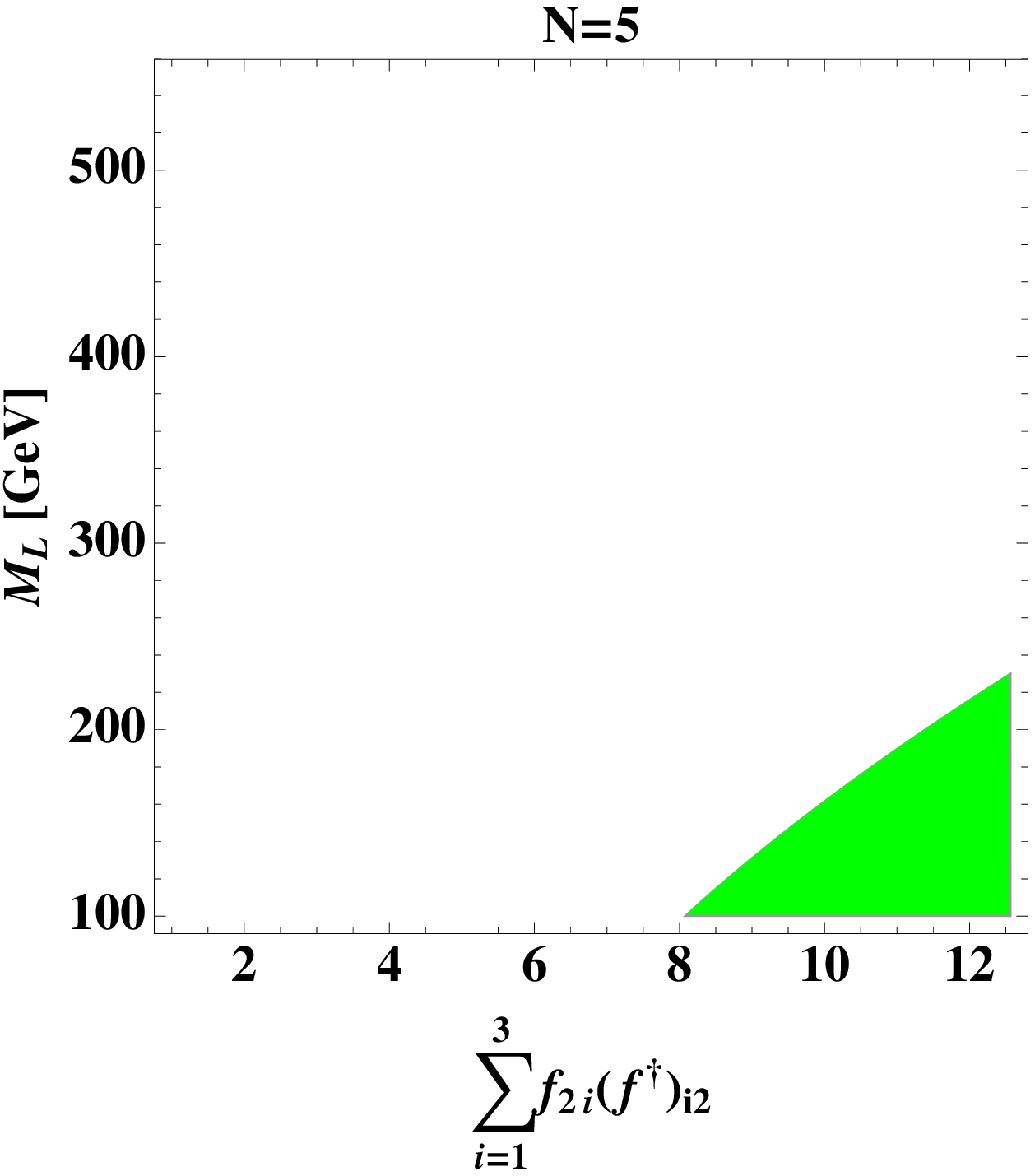}
\includegraphics[width=60mm]{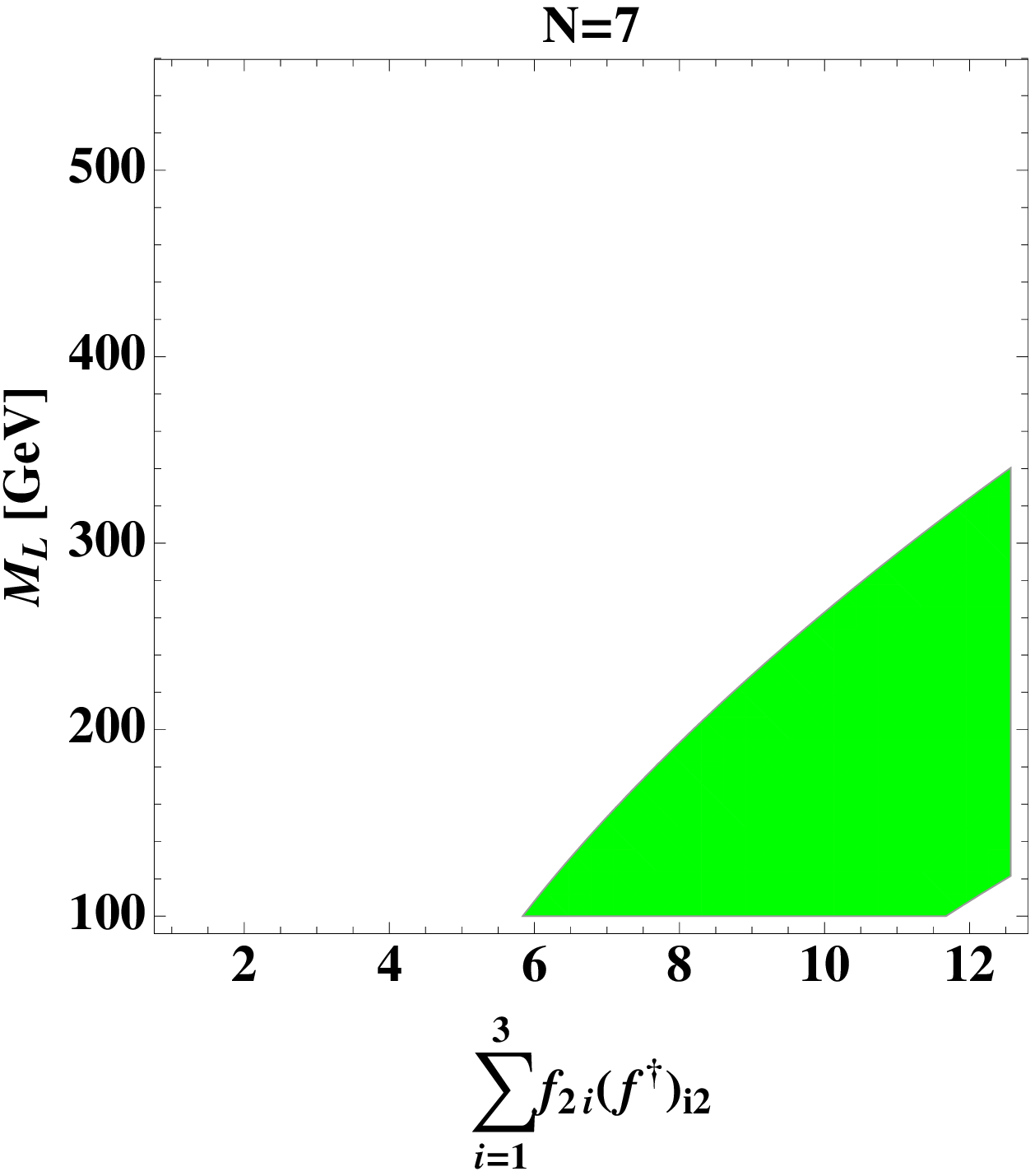}
\includegraphics[width=60mm]{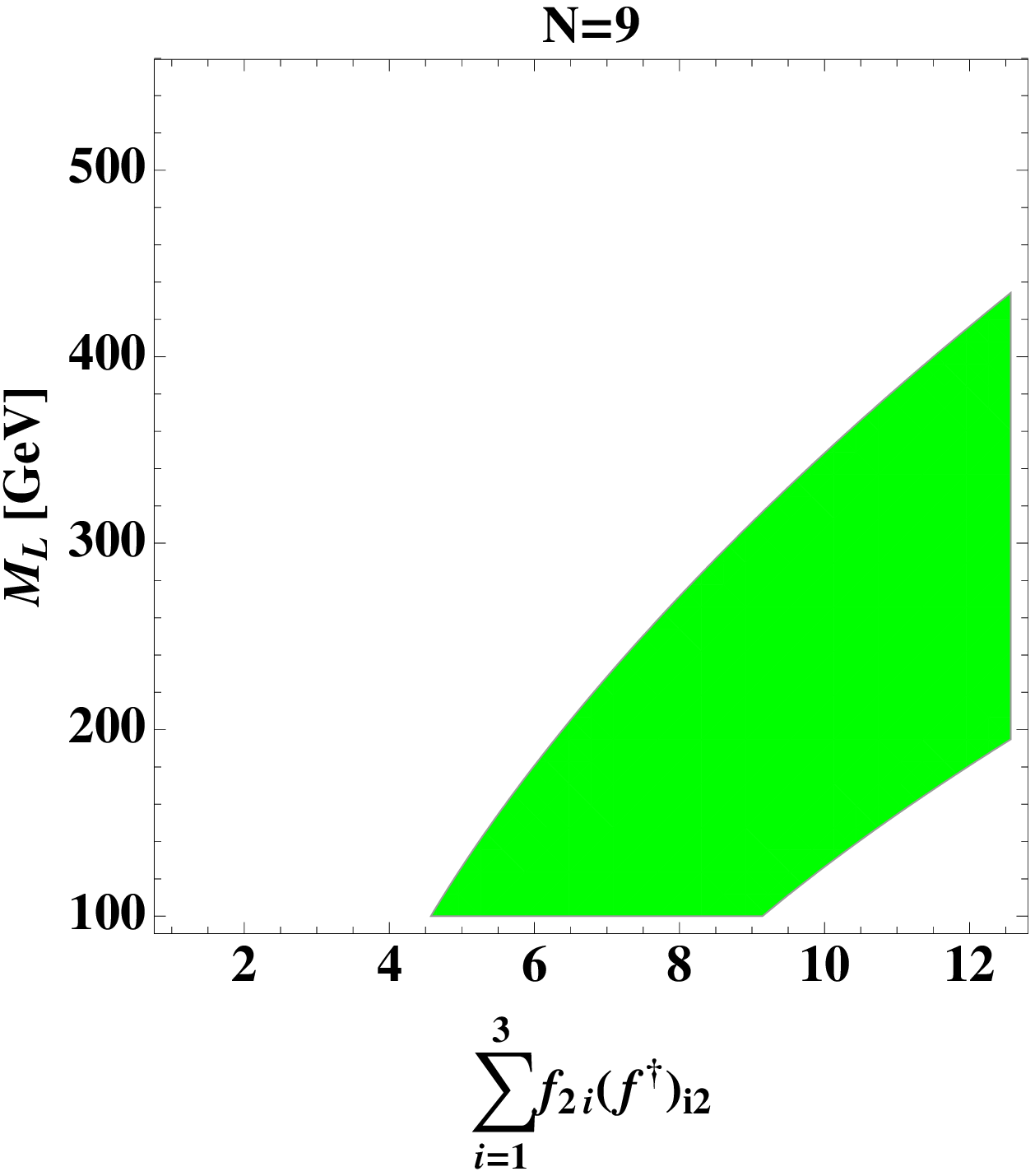}
\includegraphics[width=60mm]{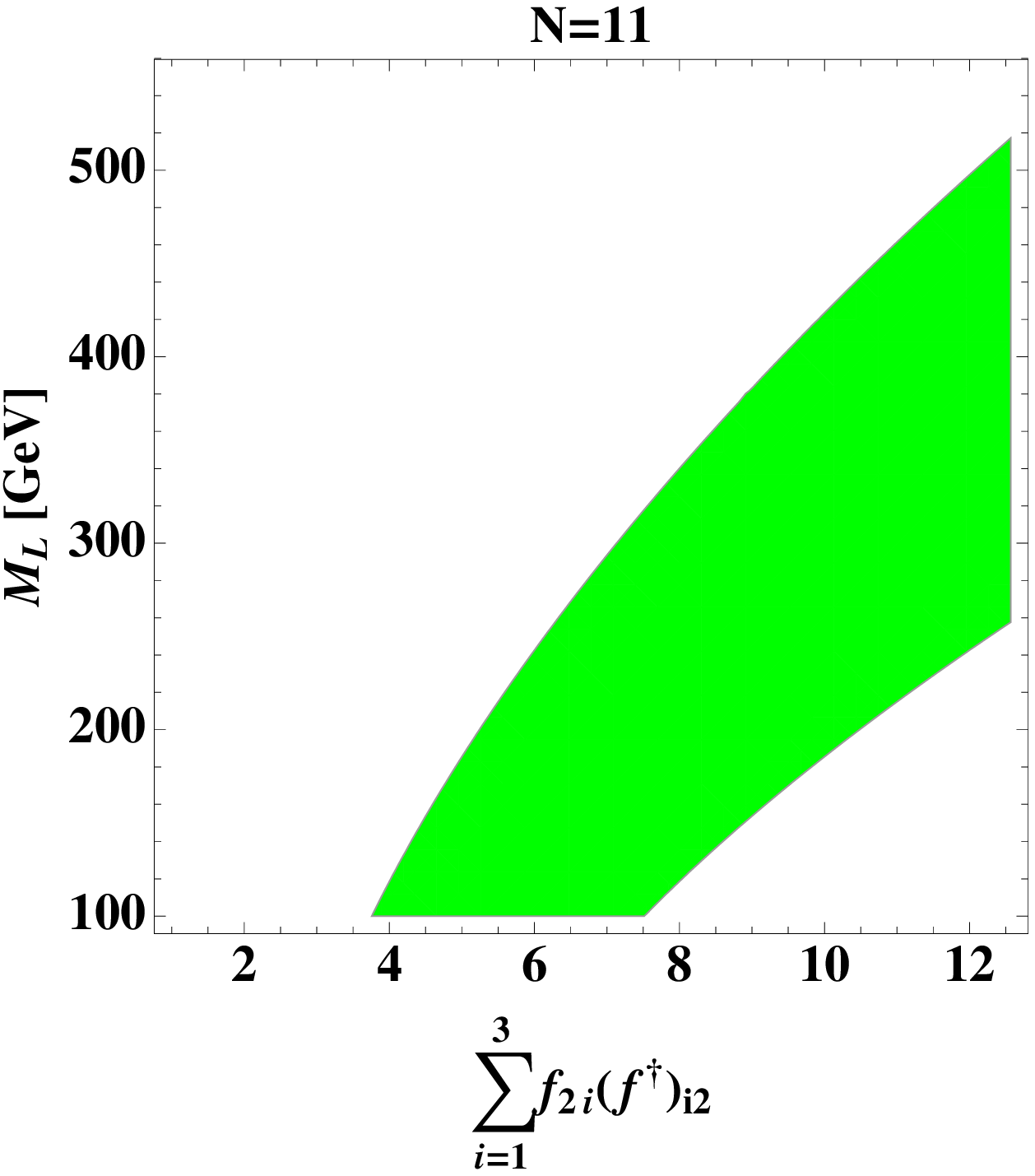}
\caption{The region plot in terms of $\sum_{i=1-3}f_{2i}(f^\dag)_{i2}$ and the $M_{L}$ plane for $N=5,7,9,11$ cases, where we fix 
$m_{S^{\pm q}}=380$ GeV  to expect the maximal diphoton excess. 
The green region satisfies the measured muon anomalous magnetic moment 
$2.0\times 10^{-9} \lesssim \Delta a_\mu\lesssim 4.0\times 10^{-9}$. 
 Notice here that $N=3$ does not have an allowed region within this range. }
\label{fig:muong-2}
\end{center}\end{figure}
%%%
%%%
Now let us turn to the muon anomalous magnetic moment $(g-2)_\mu$ within our model. 
This quantity has been measured at Brookhaven National Laboratory, and 
%The current average of the experimental results is given by~\cite{bennett}
%\begin{align}a^{\rm exp}_{\mu}=11 659 208.0(6.3)\times 10^{-10}. \notag\end{align}
%It has been well known 
there is some discrepancy between the experimental data and the prediction in the SM. 
The difference $\Delta a_{\mu}\equiv a^{\rm exp}_{\mu}-a^{\rm SM}_{\mu}$
is calculated in Refs~\cite{discrepancy1, discrepancy2} as 
\begin{align}
\Delta a_{\mu}=(29.0 \pm 9.0)\times 10^{-10},\
\Delta a_{\mu}=(33.5 \pm 8.2)\times 10^{-10}. \label{dev12}
\end{align}
These results  correspond to $3.2\sigma$ and $4.1\sigma$ deviations, respectively. 
%%%%

In our model, the muon $(g-2)_\mu$ is given by
\begin{align}
&\Delta a_\mu\approx \frac{m_\mu^2  \sum_{i=1,3} f_{2i} (f^\dag)_{i2} }{(4\pi)^2}
\biggl[ q F(E^{\pm(q+1)},S^{\pm q} ) + (q+1) F(S^{\pm q},E^{\pm(q+1)}) \biggr],\label{eq:g-2}
\\
%%%
&F(x,y)\approx \frac{
2 m_x^6 +3 m_x^4 m_y^2 - 6 m_x^2 m_y^4 +m_y^6 + 6 m_x^4 m_y^2 \ln\left[\frac{m_y^2}{m_x^2}\right]}
{12(m_x^2- m_y^2)^4}
\label{damu},
\end{align}
where we have taken the flavor universal masses for the exotic charged leptons  
for simplicity, {\it i.e.}, $M_L=M_{L_i}$.   In Fig.~\ref{fig:muong-2}, we plot the region plot 
in terms of $\sum_{i=1-3}f_{2i}(f^\dag)_{i2}$ and the $M_{L}$ plane for $N=5,7,9,11$ cases, 
where we fix $m_{S^{\pm q}}=380$ GeV to expect the maximal diphoton excess. 
The green region satisfies the measured muon anomalous magnetic moment 
$2.0\times 10^{-9} \lesssim \Delta a_\mu\lesssim 4.0\times 10^{-9}$.  
Notice here that there is no allowed parameter region for $N=3$ that can explain the deficit 
of the $a_\mu$.  Therefore, we will not discuss the case of $N=3$ in the following analysis.
%this range. 
Figure~\ref{fig:muong-2} clearly suggests that the larger value of $N$ is in favor of 
the sizable muon anomalous magnetic moment.

%\textcolor{red}{
It is worthwhile to mention the charged lepton flavor violating (CLFV) processes that are 
always induced in generating the muon anomalous magnetic moment.
In our case, CLFVs are generated from the term proportional to the Yukawa couplings $f$ 
at the one-loop level, and  the couplings or masses related to exotic fermions or bosons are 
constrained. The stringent bound is given by the $\mu\to e\gamma $ process with a penguin diagram~\cite{Adam:2013mnn}. 
However, once we 
%{\color{red}}
take  $f$ to be diagonal, such CLFVs can simply be evaded.~\footnote{
Since $g_L$ and $g_R$ are the only sources to change the flavor structure and have no 
direct interactions among SM fields,  the next leading order to the CLFVs can be induced 
at the four-loop level. Thus we expect that  the constraints are very weak.}
Even in this case,  the neutrino flavor mixings are expected to be induced via another set of Yukawa 
couplings $g_L$ and $g_R$. Hence we can retain the consistency of the CLFV constraints without 
conflict between the neutrino oscillation data and the muon anomalous magnetic moment.
%{\color{red} IS THIS A UNIQUE PROPERTY OF OUR MODEL ???}
%}

\section{DM phenomenology}

\subsection{General remarks}
%\subsection{Muon anomalous magnetic moment}
%\subsection{DM phenomenology}
%{\it Dark matter}:\\
In our model, there are two DM candidates:  a fermionic DM $\Psi_1$ and a bosonic DM $S_I$.  
%which are denoted as $\Psi_1$ and $S_I$, respectively.
Let us make some remarks for each case in the following.  
Hereafter, we shall denote either DM as $X$, and  assume that the DM pair annihilation 
into a pair of $Z'$ bosons is dominant  for simplicity. 
In this case, the elastic spin (in)dependent 
scattering is negligible, if there is no mixing between the dark  gauge boson $Z'$ and 
the SM gauge boson $Z$. 
Therefore, we can easily evade the constraint for a direct 
detection search such as LUX~\cite{Akerib:2013tjd}. 
%%%
As for the bosonic DM case especially, the constraint from direct detection can be evaded 
by having enough mass difference between the DM and its partner (the real part of the neutral 
scalar) from the $\mu$-term in Eq. (II.2) even if such a mixing cannot be negligible. 
This is because such DM always interacts with a vector boson $Z^{'}$ inelastically in the 
local $Z_2$ DM model~\cite{Baek:2014kna}. 
%Also notice here that the DM property is changed by the above decaying terms depending on 
%$N$. However, we assume that all of them are expected to be tiny because they also contribute to %the muon anomalous magnetic moment with {\it negative sign}. Thus we work the fermion DM %analysis in the first  framework of Tab.~\ref{tab:1}.

Next, we assume that all the charged scalars related to the diphoton decay are expected to 
have masses $\approx$ 380 GeV in order to enhance the 750 GeV diphoton excess, 
as we will discuss in Sec. V.   Thus the mass of DM is assumed to be  less than 380 GeV 
to make these charged scalars decay appropriately. Considering also that the mass of DM should 
be greater than the mass of $Z'$ to annihilate, we have to work on the following mass range 
for DM:  
\begin{align}
m_{Z'}\lesssim M_X \lesssim 380\ {\rm GeV}.
\end{align}

\subsection{The case of  fermion DM ($\Psi_1$)}
%{\it Case 1.  Fermion DM}\\
%%%%%%%%%%%%%%%
\begin{figure}[tb]
\begin{center}
\includegraphics[width=13cm]{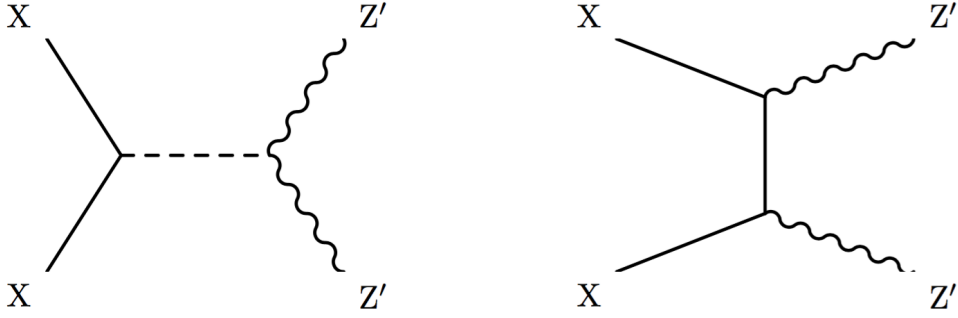}
\caption{Dominant annihilation processes for the fermionic DM.}
\label{fig:dm-ann-f}
\end{center}\end{figure}
%%%%%%%%%%%%%%%
First of all, assuming the lightest neutral particle of $\Psi_1$ as our fermion DM candidate which is denoted by $X$, 
we analyze the observed relic density $\Omega h^2\approx0.12$~\cite{Ade:2013zuv}.
The relevant interacting Lagrangian is 
\begin{align}
{\cal L}&= -(x g_X ) \bar X\gamma_\mu\gamma_5 X Z'^\mu + \frac{y_N}{4} (-s_\alpha h_{\rm SM}+c_\alpha H)\bar X(1-c_{2\theta_N})X
\nn\\&
+ {2v'( x g_X )^2}{} Z'^\mu Z'_\mu (-s_\alpha h_{\rm SM}+c_\alpha H),
\end{align}
where we have used the Majorana property of $\Psi_1$, namely $\bar X\gamma_\mu X=0$, 
in the first term.   In the following analysis, we shall take $x=1$ for simplicity.
With these interactions, we calculate the annihilation process $X X \to Z' Z'$ in 
Fig.~\ref{fig:dm-ann-f}. Then the squared spin averaged amplitude for the process is given by 
\begin{align}
|\overline{\cal M}|^2&= \frac{g_H^4}4
\left(g_{\mu,a}-\frac{k_{2\mu} k_{2a}}{m_{Z'}^2}\right) \left(g_{\nu,b}-\frac{k_{2\nu} k_{2b}}{m_{Z'}^2}\right)
\nn\\&
{\rm Tr}\left[  (\dsl p_2- M_X) 
\left(
-4 M_X(1-c_{2\theta_N}\gamma_5) g^{\mu,\nu} \left(\frac{s^2_\alpha}{s-m^2_{h_{\rm SM}}} + \frac{c^2_\alpha}{s-m^2_H} \right)
\right.\right.\nn\\
&\left.\left.
+ \gamma^\mu \gamma_5  \left(\frac{-\dsl p_1+\dsl k_1+M_X}{t-M_X^2} +\frac{-\dsl p_1+\dsl k_2+M_X}{u-M_X^2} \right)\gamma^\nu \gamma_5\right) (\dsl p_1+ M_X) \right.\nn\\
& \left.
%%%
\left(
-4 M_X(1-c_{2\theta_N}\gamma_5) g^{b,a} \left(\frac{s^2_\alpha}{s-m^2_{h_{\rm SM}}} + \frac{c^2_\alpha}{s-m^2_H} \right)
\right.\right.\nn\\
&\left.\left.
+\gamma_5 \gamma^b   \left(\frac{-\dsl p_1+\dsl k_1+M_X}{t-M_X^2} +\frac{-\dsl p_1+\dsl k_2+M_X}{u-M_X^2} \right)\gamma_5\gamma^a \right)
\right],
\end{align}
where $s,t,u$ are Mandelstam variables,
$p_1, p_2$ is the DM initial state of momentum, and $k_1, k_2$ is the $Z'$ final state of momentum.
Then the annihilation cross section is computed by 
\begin{align}
&\sigma v_{\rm rel}\approx \frac{1}{32\pi s}  \sqrt{1-\frac{4 m^2_{Z'}}{s}}
\int_0^{\pi} d\theta\sin\theta %\int_0^{\pi}d\phi 
|\overline{\cal M}|^2,  \label{eq:ann-fer}
\end{align}
and it can be expanded in terms of the relative velocity $v_{\rm rel}^2$ as
\begin{align}
\sigma v_{\rm rel} \approx a_{\rm eff} + b_{\rm eff} v_{\rm rel}^2 +{\cal O} (v_{\rm rel}^4),
\end{align}
where we take up to the $P$-wave contribution to our analysis.
%
%each of the $s$-wave ($a_{\rm eff}$) and  the $p$-wave ($b_{\rm eff}$), which can be obtained from the non-relativistic cross section of $2 X\to  2 Z'$ with $t$- and $u$-channel by expanding in terms of the relative velocity of $v_{\rm rel}$, is given by
%\begin{align}a_{\rm eff}&= \frac{(g_H x)^4 (M^6_X -M^2_X m_{Z'}^2)}{\pi m_{Z'}^4 (m_{Z'}^2 - 2M_X^2)^2},\\
%%%b_{\rm eff}&= \frac{(g_H x)^4 M_X^2 (40 M_X^8 - 68 M^6_X m_{Z'}^2 +112 M_X^4  m_{Z'}^4 - 77 M^2_X m_{Z'}^6+23  m_{Z'}^8)}{24\pi (m_{Z'}^3 - 2M_X^2 m_{Z'})^4}\sqrt{1-\frac{ m_{Z'}^2}{M_X^2}},\end{align}
%The $p$-wave is also written, all the contributions, therefore, $2X\to 2h_{\rm SM}$, and $2 X\to  2 Z'$, 
% but we do not explicitly write down because of its complication.
 %%%
Thus the relic density is given by~\cite{Gondolo:1990dk}
\begin{align}
\Omega h^2\approx \frac{1.07\times10^9 x_f^2}
{g^{1/2}_* M_{\rm pl}[{\rm GeV}] 
(a_{\rm eff} x_f + 3b_{\rm eff})},
%\int_{x_f}^\infty \left(\frac{a_{\rm eff}}{x^2}+6\frac{b_{\rm eff}}{x^3} \right)},
\label{eq:relic}
\end{align}
where $g_*\approx 100$ is the total number of effective relativistic degrees of freedom at the time of freeze-out,
$M_{\rm pl}=1.22\times 10^{19}[{\rm GeV}] $ is the Planck mass, and $x_f\approx25$.
The observed relic density reported by Planck suggests that $\Omega h^2\approx 0.12$~\cite{Ade:2013zuv}.

\begin{figure}[tb]
\begin{center}
\includegraphics[width=13cm]{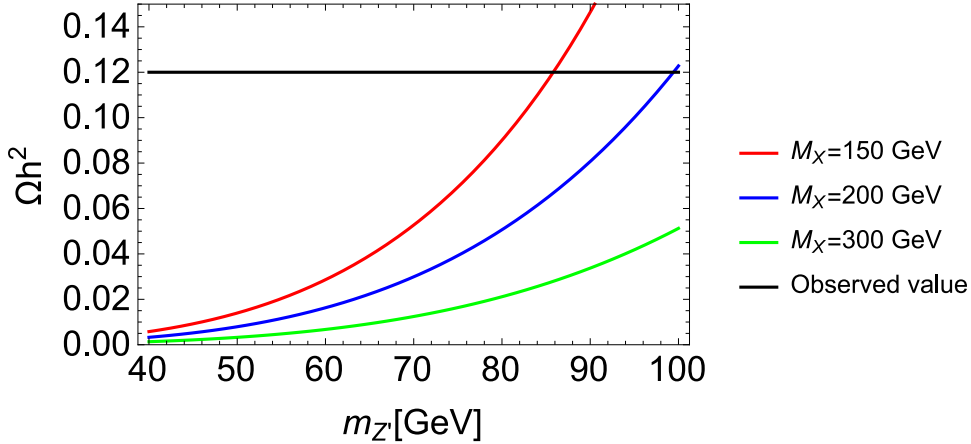}
\caption{Thermal relic density of fermionic DM as a function of the dark gauge boson 
mass for three different values of DM mass.}
\label{fig:ferDM}
\end{center}\end{figure}

In Fig.~\ref{fig:ferDM}, we show the thermal relic density of fermionic 
DM as a function of  the dark gauge boson mass $m_{Z'}$. 
% as a function of hidden gauge coupling.
We fix other parameters as follows: 
\begin{eqnarray}
g_X =0.1, \ \ m_H=750\ {\rm GeV} . 
\end{eqnarray}
The three lines correspond to three DM masses: the red, blue, and green curves represent 
DM masses equal to 150 GeV,  200 GeV, and 300 GeV,  respectively.

\subsection{The case of bosonic DM ($S_I$)}
%{\it Case 2.  Boson DM}\\
%%%%%%%%%%%%%%%
\begin{figure}[tb]
\begin{center}
\includegraphics[width=13cm]{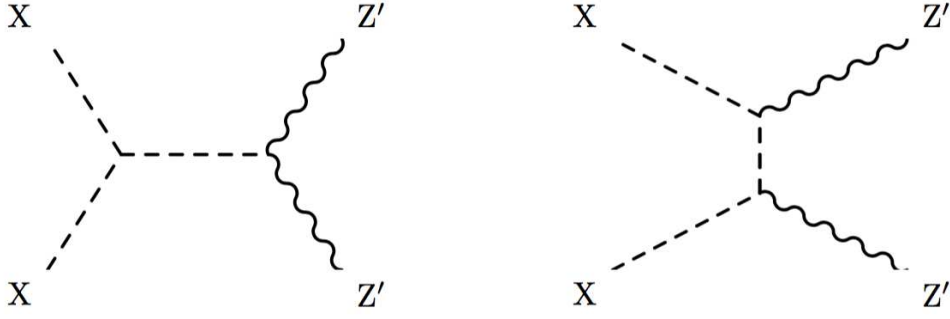}
\caption{Dominant annihilation processes for the bosonic DM.}
\label{fig:dm-ann-b}
\end{center}\end{figure}
%%%%%%%%%%%%%%%
Next, we consider the bosonic DM, assuming $S_I$ to be the DM candidate $X$.
The relevant interacting Lagrangian to estimate the relic density is 
\begin{align}
{\cal L}= -(x g_X)(S_R\partial_\mu X-X\partial_\mu S_R) Z'^\mu + \frac12 (x g_X)^2 Z'_\mu Z'^\mu X^2
-\frac{\mu_{2XH}}{4} s_I^2 H -\frac{\mu_{2Xh}}{4} s_I^2 h_{\rm SM} ,
\end{align}
where we define 
\begin{eqnarray}
\mu_{2XH} & \equiv & \lambda_{S\Phi}vs_\alpha +\lambda_{S\varphi} v' c_\alpha
\label{eq:mu1}
\\
\mu_{2Xh} & \equiv & \lambda_{S\Phi}v c_\alpha -\lambda_{S\varphi} v' s_\alpha .
\label{eq:mu2}
\end{eqnarray}
In the following analysis, we shall take $x=1$ for simplicity.
With these interactions, we calculate the annihilation process $X X \to Z' Z'$ in 
Fig.~\ref{fig:dm-ann-b}. Then the squared spin averaged amplitude for the process is given by 
\begin{align}
&|\overline{\cal M}|^2= {g_H^4}
\left(g_{\mu,a}-\frac{k_{2\mu} k_{2a}}{m_{Z'}^2}\right) \left(g_{\nu,b}-\frac{k_{2\nu} k_{2b}}{m_{Z'}^2}\right)\nn\\
%%%
&\left[ 
2g^{\mu\nu}\left(1-v'\left[\frac{\mu_{2Xh}s_\alpha}{s-m^2_{h_{\rm SM}}} - \frac{\mu_{2XH}c_\alpha}{s-m^2_{H}} \right] \right)%\right.\nn\\ &\left.
+(p_2 + p_1-k_1)_\mu
\left( \frac{(2p_1-k_1)_\nu}{t-m^2_{S_R}}
+\frac{(2p_2-k_1)_\nu}{u-m^2_{S_R}} \right)
\right]\nn\\
&\left[ 
2g^{ab}\left(1-v'\left[\frac{\mu_{2Xh}s_\alpha}{s-m^2_{h_{\rm SM}}} - \frac{\mu_{2XH}c_\alpha}{s-m^2_{H}} \right] \right)%\right.\nn\\ &\left.
+(p_2 + p_1-k_1)_a
\left( \frac{(2p_1-k_1)_b}{t-m^2_{S_R}}
+\frac{(2p_2-k_1)_b}{u-m^2_{S_R}} \right)
\right]
,
\end{align}
where the other process is the same as in the fermion DM case; therefore, the relic density is computed by substituting the above mass invariant squared into Eqs.~(\ref{eq:ann-fer}) and (\ref{eq:relic}).

\begin{figure}[tb]
\begin{center}
\includegraphics[width=15cm]{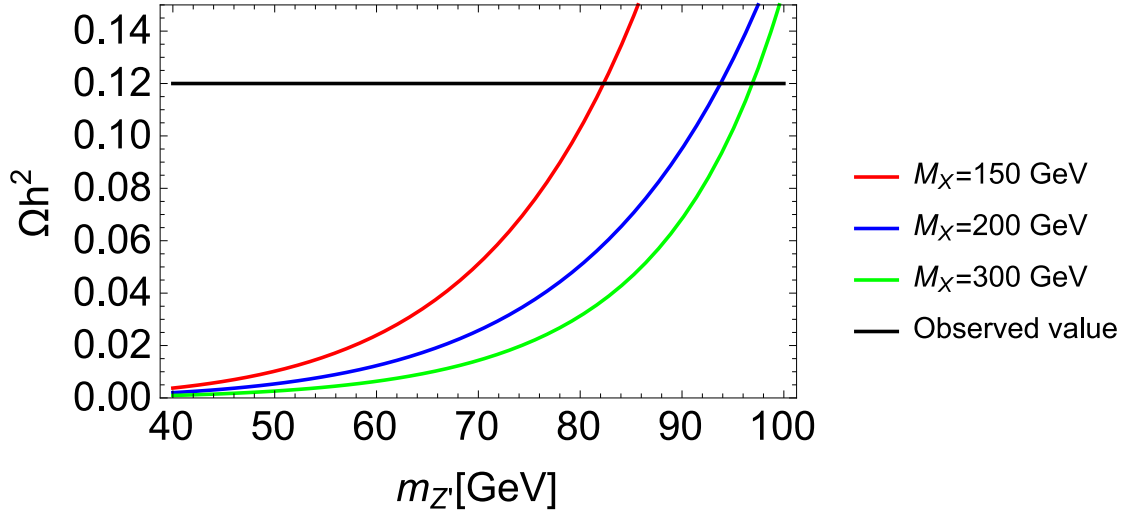}
\caption{Thermal relic density of bosonic DM as a function of the dark gauge boson 
mass for three different values of DM mass.}
\label{fig:scalarDM}
\end{center}\end{figure}

In Fig.~\ref{fig:scalarDM}, we show the DM relic density as a function 
of the dark gauge boson mass $m_{Z'}$ for the following values of the relevant 
parameters: 
\begin{eqnarray}
g_X=0.1, \ \ \sin\alpha = 0.2, \ \ \lambda_{S\Phi}= 0.1,  \ \ 
\lambda_{S\varphi}= 0.1, \ \ m_H=750\ {\rm GeV}. 
\end{eqnarray}
The three lines correspond to three DM different masses: 150 GeV (red), 200 GeV (blue), 
and  300 GeV (green). 
\section{750 GeV Diphoton excess}
In this section, we discuss how we can explain the diphoton excess at 750 GeV 
within our models for $N=5,7,9$, and 11.
The candidate of 750 GeV diphoton resonance in our model is the scalar particle $H$, which 
is a linear combination of the CP-even neutral components of $\Phi$ and $\varphi$.
In our model, the diphoton decay channel of $H$ is induced by the interactions of 
$\Phi$ and $\varphi$ with charged scalars which can be generally written as 
\begin{equation}
{\cal L} \supset \sum_{i} \left[ \lambda_{\Phi \phi_i^Q} |\Phi|^2 |\phi_i^Q|^2 + \lambda_{\varphi \phi_i^Q} | \varphi|^2 |\phi^Q_i|^2 \right], 
\end{equation}
where charged scalar fields with electric charge $Q$ are denoted by  $\phi^Q_i$. 
The charged scalar fields for $N=5,7,9,11$ are specified as 
\begin{align}
& N = 5 \, : \, \phi^Q = \{\phi^{\pm \pm}, \phi^{\pm \pm \pm}, S^{\pm \pm}, D^\pm  \}, \nonumber \\
& N = 7 \, : \, \phi^Q = \{\phi^{\pm \pm \pm}, \phi^{\pm \pm \pm \pm}, S^{\pm \pm \pm}, D^\pm, D^{\pm \pm}   \}, \nonumber \\
& N = 9 \, : \, \phi^Q = \{\phi^{\pm \pm \pm \pm}, \phi^{\pm \pm \pm \pm \pm}, S^{\pm \pm \pm \pm}, D^{\pm \pm}  \}, \nonumber \\
& N = 11 \, : \, \phi^Q = \{\phi^{\pm \pm \pm \pm \pm}, \phi^{\pm \pm \pm \pm \pm \pm}, S^{\pm \pm \pm \pm \pm}, D^\pm, D^{\pm \pm}  \}.
\end{align}
After symmetry breaking, trilinear interactions among the mass eigenstates are given by
\begin{equation}
{\cal L} \supset  \sum_i  \left[ (\lambda_{\Phi \phi_i^Q} v \cos  \alpha - \lambda_{\varphi \phi_i^Q} v' \sin  \alpha) h_{\rm SM} |\phi^Q_i|^2 + (\lambda_{\Phi \phi_i^Q} v \sin \alpha + \lambda_{\varphi \phi_i^Q} v' \cos  \alpha) H |\phi^Q_i|^2 \right],
\end{equation}
where the mixing angle $\alpha$ is given in Eq.~(\ref{eq:alpha}) . 
Here we require the contribution to $h_{\rm SM} \to \gamma \gamma$ from new charged scalars 
$\phi_i^Q$ to be suppressed by assuming $\lambda_{\Phi \phi^Q_i} v \cos  \alpha \simeq \lambda_{\varphi \phi^Q_i} v' \sin  \alpha$.  This would make our model consistent 
with the LHC data on the 125 GeV Higgs signal strengths. 
Then,  the Lagrangian involving the trilinear couplings for $H(750)$ is obtained as 
\begin{equation}
\label{eq:tri}
\sum_i \frac{\lambda_{\varphi \phi^Q_i} v'}{\cos  \alpha} H |\phi^Q_i|^2 = 
\sum_i \mu_{H \phi^Q_i} H |\phi^Q_i|^2,
\end{equation}
where $\mu_{H \phi^Q_i} = \lambda_{\varphi \phi^Q_i} v'/ \cos  \alpha$ is the trilinear coupling.

The scalar particle $H$ can be produced by gluon fusion through mixing with SM Higgs.
The cross section is given by
\begin{equation}
\sigma(gg \to H) \simeq \sin^2  \alpha \times 0.85 \ {\rm pb},
\end{equation}
at the LHC 13 TeV~\cite{Djouadi:2013uqa,Khachatryan:2014wca}.
Moreover, $H$ can be produced by photon fusion, $pp(\gamma \gamma) \to H$, in our model 
due to the sizable effective $H \gamma \gamma$ coupling by charged scalar loop contributions.
Here we adopt the estimation of the photon fusion cross section including both elastic and 
inelastic scattering in Ref.~\cite{Csaki:2016raa}: 
\begin{equation}
\label{eq:photon-fusion}
\sigma(pp (\gamma \gamma) \to H \to \gamma \gamma +X)_{\rm 13 TeV} = 10.8{\rm pb} \left( \frac{\Gamma_H}{45 {\rm GeV}}  \right) \times BR^2(H \to \gamma \gamma),
\end{equation}
where $X$ denotes any other associated final states. 
Therefore the total cross section for $pp \to H \to \gamma \gamma$ would be determined by 
\begin{equation}
\label{eq:CXgaga}
\sigma_{\gamma \gamma} = \sigma (gg \to H) BR(H \to \gamma \gamma) + 
\sigma_{\gamma-{\rm fusion}} \ , 
\end{equation}
where $\sigma_{\gamma-{\rm fusion}}$ is from Eq.~(\ref{eq:photon-fusion}).

\begin{figure}[t]
\begin{center}
\includegraphics[width=80mm]{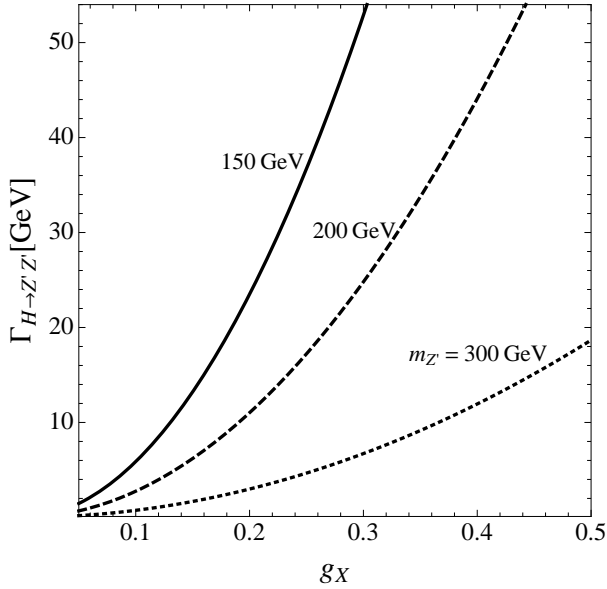} 
\caption{The partial decay width for $H \to Z' Z'$ as a function of $g_X$ for $m_{Z'} = 150$, 
200, and 300 GeV.
}   \label{fig:width}
\end{center}\end{figure}

Decays of $H$ into SM particles are induced via mixing with SM Higgs, where the dominant 
partial decay widths are 
\begin{align}
\Gamma(H \to W^+ W^-) =&  \frac{g^2 m_W^2 \sin^2  \alpha}{64 \pi m_{H}} \frac{m_{H}^4 - 4 m_{H}^2 m_W^2 + 12 m_W^2}{m_W^4} \sqrt{1- \frac{(2 m_W)^2}{m_{H}^2}}\,,  \\
\Gamma(H \to Z Z) =& \frac{1}{2} \frac{g^2 m_Z^2 \sin^2  \alpha}{64 \pi \cos^2 \theta_W m_{H}} \frac{m_{H}^4 - 4 m_{H}^2 m_Z^2 + 12 m_Z^2}{m_Z^4} \sqrt{1- \frac{(2 m_Z)^2}{m^2_{H}} }\,,  \\
\Gamma(H \to t \bar t) = & \frac{3 m_t^2 \sin^2 \alpha}{8 \pi v^2} m_H \sqrt{1 - \frac{4 m_{t}^2}{m_H^2}}.
\end{align}
We note that partial decay widths for other SM fermion final states are subdominant.
The diphoton decay $H \to \gamma \gamma $ is generated dominantly  by charged scalar 
loops within our model, whose  partial decay width is given by~\cite{Gunion:1989we}
\begin{equation}
\Gamma_{H \to \gamma \gamma} \simeq \frac{\alpha_{\rm em}^2 m_H^3}{256 \pi^3} \left| \sum_{\phi^{Q}_i} Q_i^2  \frac{\mu_{H \phi^Q_i} }{2 m_{\phi^Q_i}^2} A_0 (\tau_{\phi^Q_i}) \right|^2,
\end{equation}
where $\alpha_{\rm em}\approx1/137$ is the fine structure constant, $A_0 (x) = -x^2[x^{-1} - [\sin^{-1} (1/\sqrt{x})]^2]$ and 
$\tau_{\phi^Q_i} = 4 m_{\phi^Q_i}^2/m_H^2$ and we omit SM particle contributions 
since they are small compared with charged scalar contributions. 
Here we note that the $H \to Z\gamma$ mode is also induced at the one-loop level. 
Since it is subdominant contribution, we shall omit the explicit formula for the partial width.
When $2 m_{Z'} < m_H$ the decay channel $H \to Z' Z'$ opens. Its partial decay width is 
given by 
\begin{equation}
\Gamma(H \to Z' Z') = \frac{g_H^2 m_{Z'}^2}{8 \pi m_H}  \frac{m_H^4 - 4 m_H^2 m_{Z'}^2 
+ 12 m_{Z'}^4}{m_{Z'}^4} \sqrt{1 - \frac{4 m_{Z'}^2}{m_H^2}}\,.
\end{equation}
The partial decay width for $H \to Z' Z'$ is shown in Fig.~\ref{fig:width} as a function of 
$g_X$ for several values of $m_{Z'}$.
The decay modes $H \to S_{I(R)} S_{I(R)}$ and $H \to N_{1(2)} N_{1(2)}$ are also possible when they are kinematically allowed; 
partial decay widths of these modes are obtained as 
\begin{align}
\Gamma(H \to S_{I(R)} S_{I(R)}) & = \frac{\mu_{2XH}^2 }{16 \pi m_H} \sqrt{1- \frac{4 m_{S_{I(R)}}^2}{m_H^2}} \  , \\
\Gamma(H \to N_{1(2)} N_{1(2)}) & = \frac{y_{N_{L(R)}}^2 c_{\theta_N}^2 + y_{N_{R(L)}}^2 s_{\theta_N}^2}{64 \pi} m_H \left( 1- \frac{2 m_{N_{1(2)}}^2}{m_H^2} \right) \sqrt{1 - \frac{4 m_{N_{1(2)}}}{m_H^2}} \ ,
\end{align}  
where $\mu_{2XH}$ is defined in Eq.~\ref{eq:mu2}.
We note that these partial decay widths are subdominant compared to the $H \to Z' Z'$ mode.
%%%%%%%%%%%%%%%%%%%
%%%%%%%%%%%%%%%%%%%

The constraint from 8 TeV LHC data for diphoton searches should be taken into account, 
since $BR(H \to \gamma \gamma)$ can be sizable in our model. 
We take the following  as the constraint: % is given by 
\begin{equation}
\sigma_{\gamma \gamma}^{\rm 8 TeV} \equiv \sigma (gg \to H)^{\rm 8 TeV} BR(H \to \gamma \gamma) + \sigma_{\gamma-{\rm fusion}}^{\rm 8 TeV} < 1.5 \ {\rm fb} \ .
\end{equation}
The ratio of a 13 TeV cross section and an 8 TeV cross section for gluon fusion is estimated as $\sigma(gg \to H)^{\rm 13 TeV}/\sigma(gg \to H)^{\rm 8 TeV} \simeq 5$~\cite{Franceschini:2015kwy}. 
For the photon fusion process, we write the ratio as 
$\sigma_{\gamma-{\rm fusion}}^{\rm 13 TeV}/\sigma_{\gamma-{\rm fusion}}^{\rm 8 TeV} \equiv R_{\gamma \gamma}$. 
Here $R_{\gamma \gamma}$ is estimated to be $\sim 2$~
\cite{Csaki:2016raa} but the uncertainty is large, so it can be a larger value~
\cite{Fichet:2015vvy,Csaki:2016raa,Harland-Lang:2016qjy}. 
In our analysis, we investigate 
the constraint using $R_{\gamma \gamma}=2$ and 4 as reference values.

%%%%%%%%%%%%%%%%%%%
\begin{figure}[!hptb]
\begin{center}
\includegraphics[width=60mm]{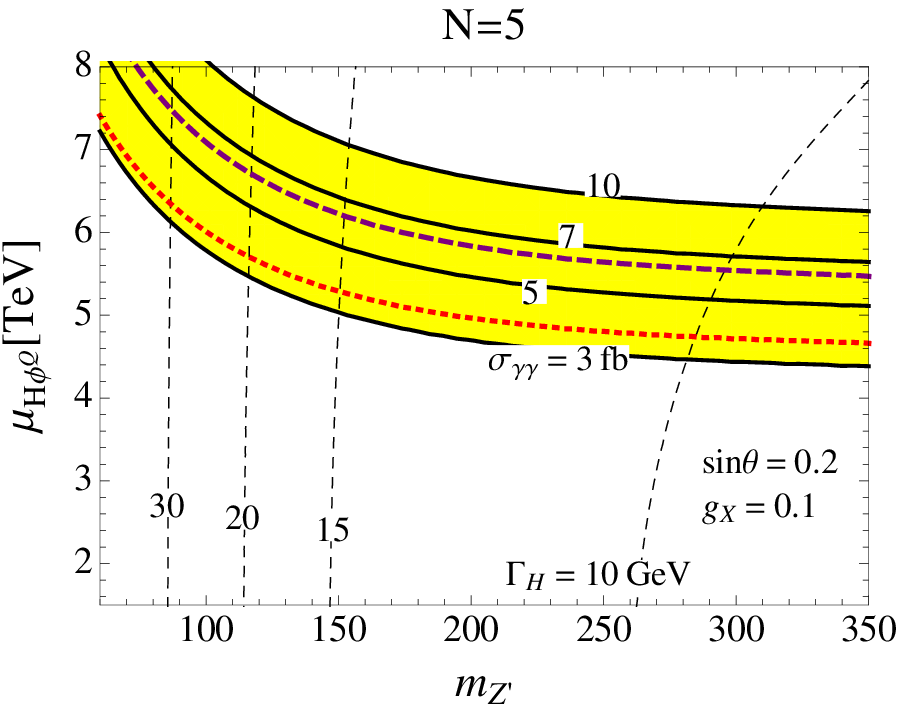} \qquad
\includegraphics[width=60mm]{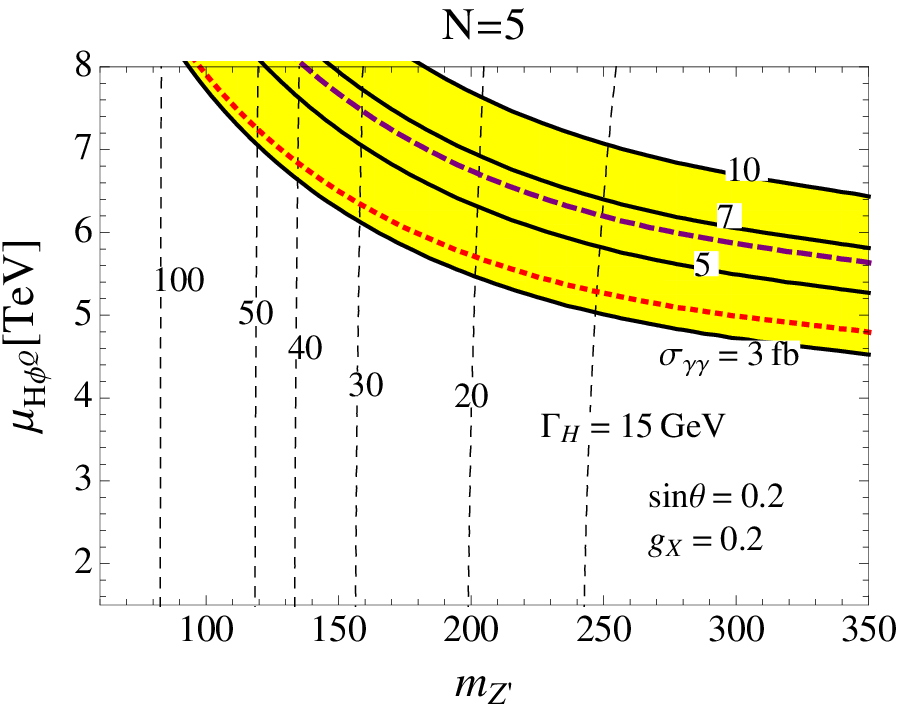}
\includegraphics[width=60mm]{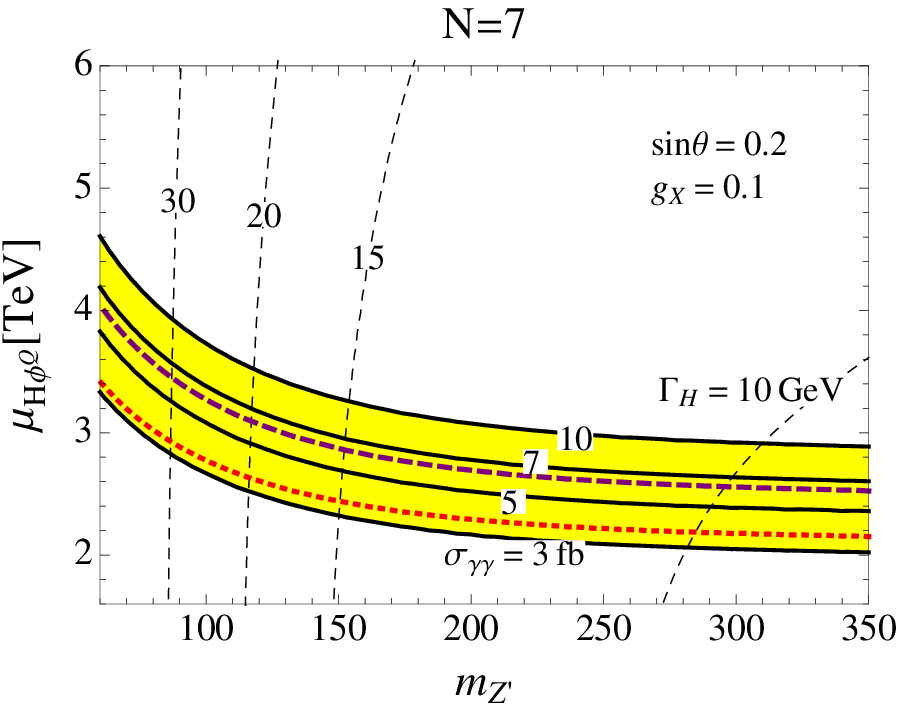} \qquad
\includegraphics[width=60mm]{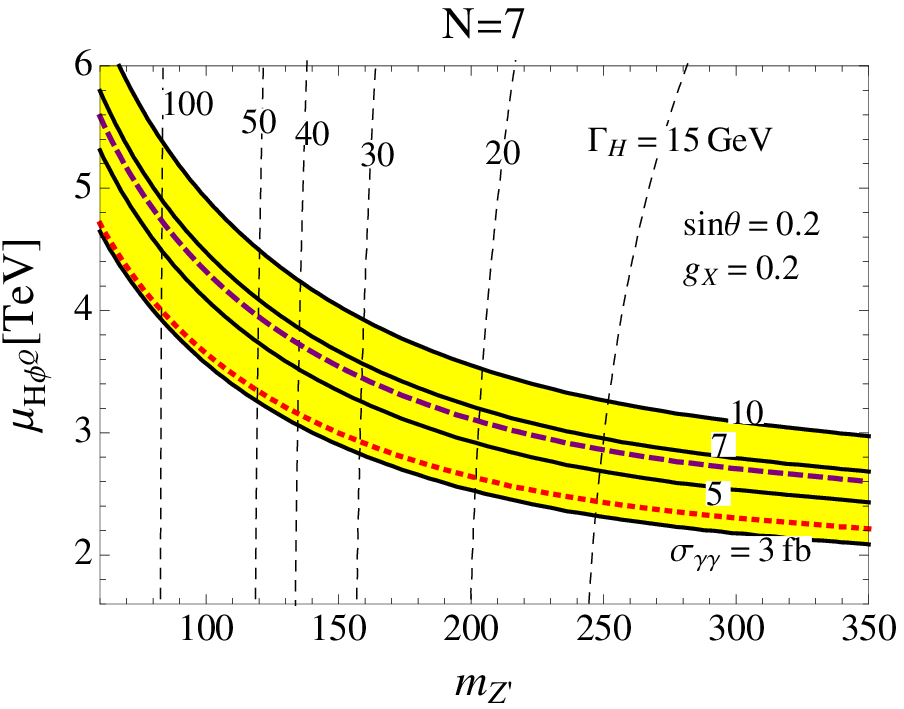}
\includegraphics[width=60mm]{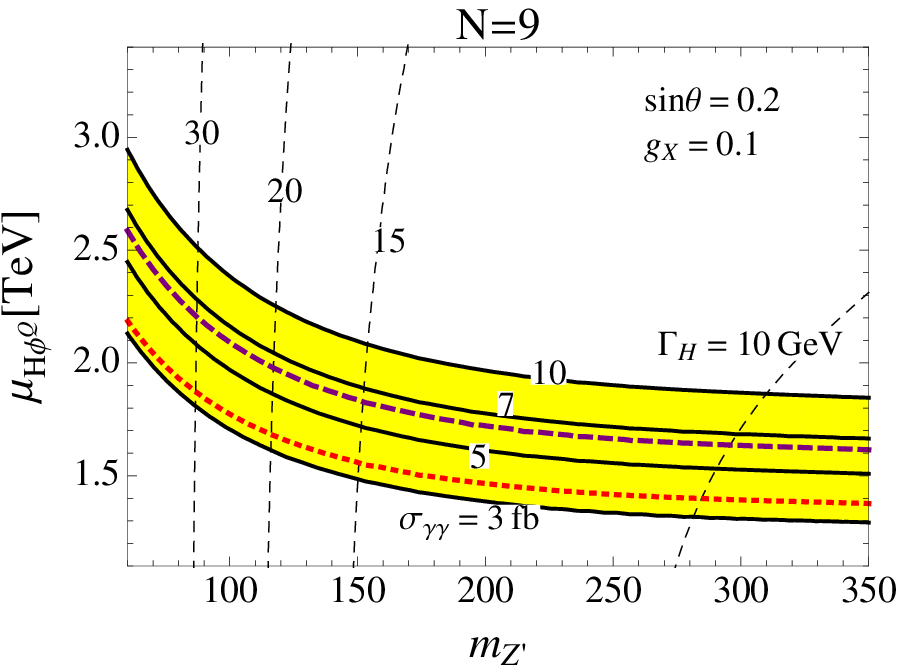} \qquad
\includegraphics[width=60mm]{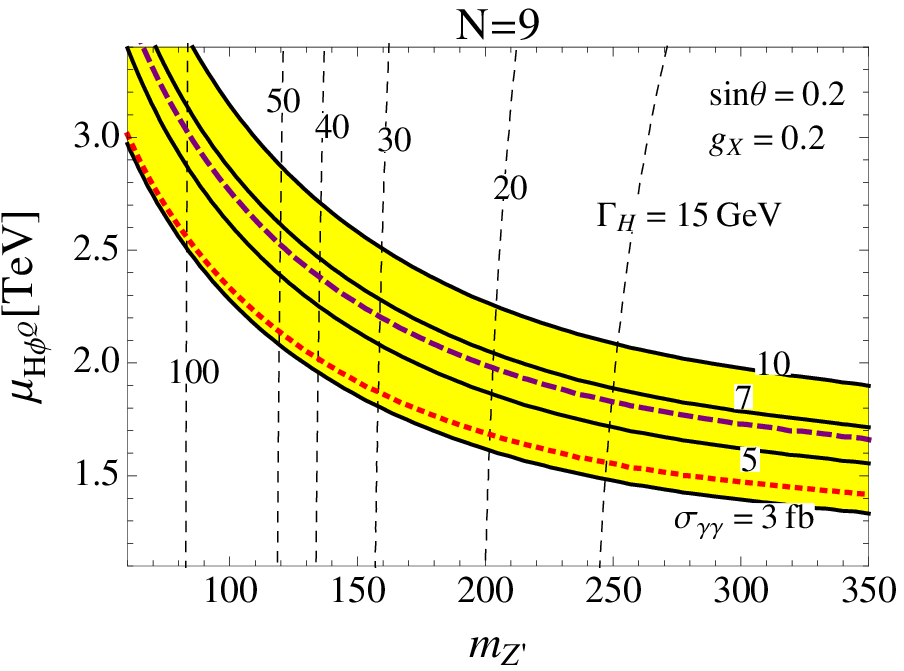}
\includegraphics[width=60mm]{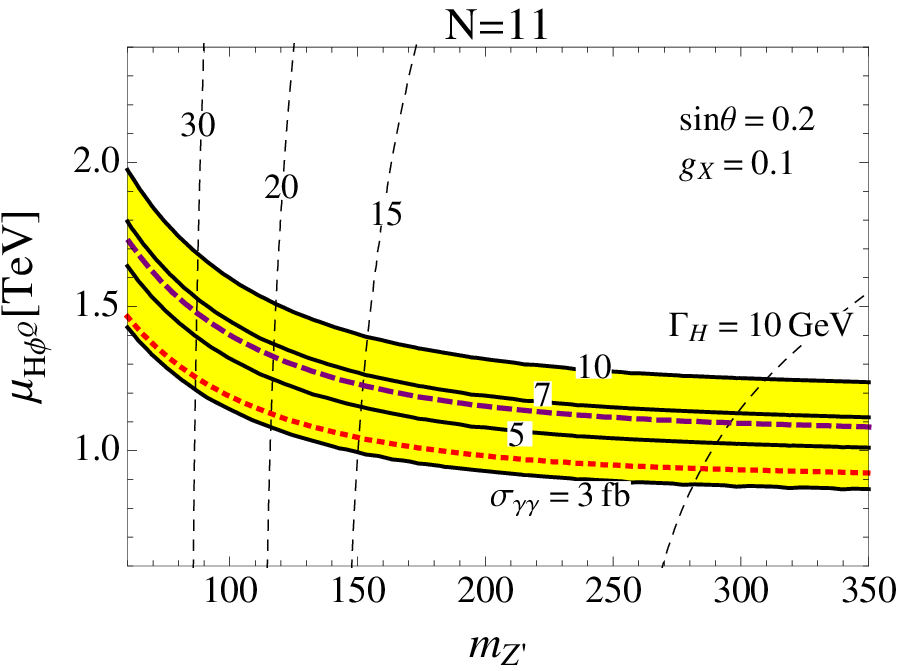} \qquad
\includegraphics[width=60mm]{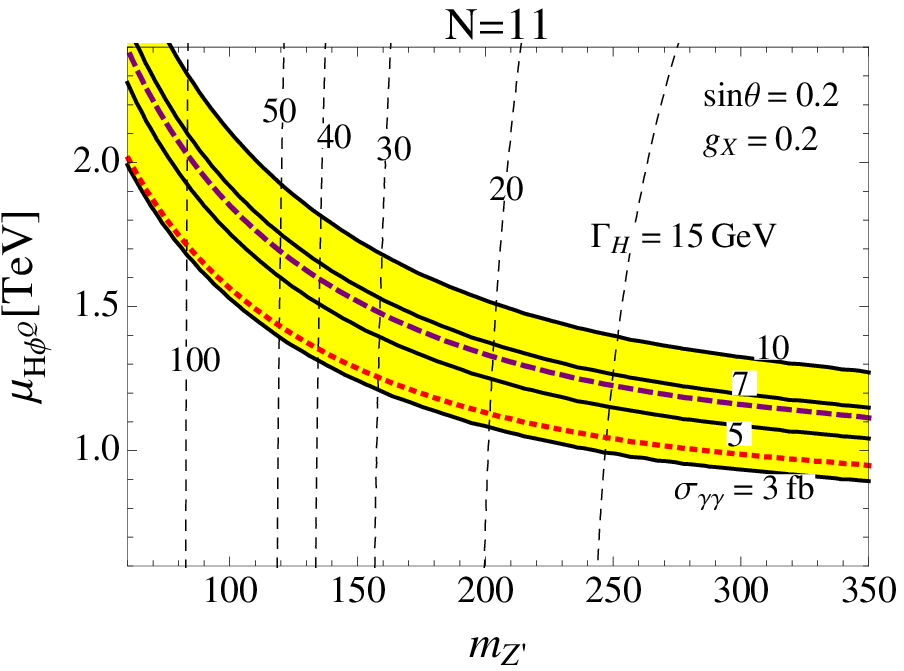}
\caption{The contours of $\sigma (gg \to H) BR(\phi \to \gamma \gamma)$ (in units of fb) and total width $\Gamma_H$ (in units of GeV) in $m_{Z'} - \mu_{H \phi^Q}$ plane for $N= 5,7,9,11$.
All the charged scalar masses are taken to be 380 GeV. The purple dashed and the red dotted lines indicate the constraints from diphoton searches at 8 TeV for $R_{\gamma \gamma} =2$ and $4$ where the region above the lines is excluded.
}   \label{fig:diphoton1}
\end{center}\end{figure}
%%%%%%%%%%%%%%%%%%%
Finally, we  estimate $\sigma_{\gamma \gamma}$ for the cases of $N=5,7,9$, and $11$ where 
we assume the couplings $\mu_{H \phi^Q_i}$ take the same value for all charged scalars for simplicity. 
Also, the mass of the charged scalar is set to be $m_{\phi^Q_i} = 380$ GeV to enhance loop function inside the diphoton decay width.
We show the contours of $\sigma_{\gamma \gamma}^{\rm}$ and $\Gamma_H$ in Fig.~\ref{fig:diphoton1} by solid and dashed lines,
 respectively, for $N= \{ 5,7, 9,11\}$, 
and the constraints from diphoton searches at 8 TeV for $R_{\gamma \gamma} =2$ and $4$ are indicated by the purple dashed and the red dotted lines where the region above the lines is excluded.
We find that contribution from the photon fusion process is dominant in the parameter region, explaining the diphoton excess.
Note here that the contribution from gluon fusion is required to avoid the constraint from an 8 TeV diphoton search for $R_{\gamma \gamma} \simeq 2$, 
so we adopt $\sin \theta =0.2$, which is allowed by the constraint of the SM Higgs mixing 
angle~\cite{hdecay, Chpoi:2013wga,Cheung:2015dta,Cheung:2015cug}. 
Then we can obtain $\sim 3$ fb cross section for all $R_{\gamma \gamma }$ =2 and $\gtrsim 5$ fb for $R_{\gamma \gamma} =4$.
%The trilinear coupling can be $O(1)$ TeV for $N \geq 5$, which is safe from perturbative unitarity. 
We also find that the total decay width $\Gamma_H$ is $O (10)$ GeV and becomes larger for larger $g_X$ due to contribution from the $H \to Z'Z'$ channel.

 Here we comment on collider phenomenology of the $Z'$ boson. The $Z'$ couples to SM particles via $Z$-$Z'$ mixing which is induced from kinetic mixing between $U(1)_Y$ and $U(1)_X$ gauge fields, $(\xi/4) F^{\mu \nu}_Y F_{X \mu \nu}$. 
 %Since we assume the kinetic mixing is very small, cross section for direct $Z'$ production process is also small. 
 For $m_{Z'} \sim 100$ GeV, the kinetic mixing parameter is limited as $\xi \lesssim 10^{-3}-10^{-2}$ experimentally~\cite{Andreas:2012mt, Jaeckel:2012yz}.   
 With this tiny kinetic mixing parameter, we have a very small cross section for $Z'$ production via the Drell-Yan process.
 Thus, $Z'$ will be dominantly produced through the process $pp \to  H \to Z' Z'$ without suppression by kinetic mixing.  The produced $Z'$ then decays such that $Z' \to jj, \ell^+ \ell^-$, etc. through the mixing effect. 
 %The $Z'$ production process can be tested at the LHC 13 TeV since $pp \to {\color{red} H}$ production cross section is order of $\sigma \sim \sin^2 \alpha \times 0.85 \, \text{pb}  \sim 34 \, \text{fb}$ with $\sin \alpha = 0.2$ and $BR(\phi \to Z' Z') \sim 1$ when it is kinematically allowed.
 From Eq.~(\ref{eq:CXgaga}), we can derive the $H$ production cross section as $\sigma (pp \to H) = \sigma(gg \to H) + \sigma_{\gamma-{\rm fusion}}/BR(H \to \gamma \gamma)$.
Then we obtain $\sigma (pp \to H) \simeq 184$ fb at the LHC 13 TeV 
with a reference parameter set $N=11$,  $\sin \alpha = 0.2$, $g_X = 0.1$, $\mu_{H\phi^Q }=1.3$ TeV, and  $m_{Z'} =100$ GeV, which can explain diphoton excess.
Thus, the $Z'$ production cross section can be sizable, since $BR(H \to Z' Z') \simeq 1$, and the LHC experiments can explore the $Z'$ production process with sufficient luminosity.
For $\xi \sim 10^{-3}$, the decay width of $Z'$ is roughly $\Gamma_{Z'}/m_{Z'} \sim O(\xi^2) \sim 10^{-6}$, which provides a lifetime of $Z'$ as $\sim 10^{-20}$s for $m_{Z'} \sim 100$ GeV. 
Then the produced $Z'$ will decay before reaching the detectors at the LHC. 
We also note that the $Z'$ production cross section for  $pp \to h_{\rm SM} \to Z'Z'$ provides a small contribution for $m_{Z'} > m_h/2$, which is preferred by the relic density of DM, since $h_{\rm SM}$ is off-shell.

Before closing this section,  let us discuss the consistency of DM relic density calculation 
with perturbative unitarity constraints on the trilinear scalar couplings, which are given by 
Eq.~(\ref{eq:tri}).   Note that the VEV of $\varphi$ is related to $g_X$ and $m_{Z'}$ as in  
Eq.~(\ref{eq:Z'mass}).  Taking $g_X = 0.1$ and $x =1$ as in Sec.III,  the value of $v'$ in our 
scenario is typically 400 to 500 GeV since $m_{Z'} \simeq 80$ to 100 GeV to explain the relic density of DM as shown in Figs.~\ref{fig:ferDM} and \ref{fig:scalarDM}.
Thus the trilinear coupling should satisfy $\mu_{H \phi^Q} \lesssim 1.8$ TeV when we require $\lambda_{H \phi^Q} \lesssim \sqrt{4 \pi}$ to satisfy perturbative unitarity safely.
Therefore, the $N=9$ and $N=11$ cases have parameter spaces satisfying the condition and explaining the diphoton excess, while the $N=5$ and $N=7$ cases require trilinear coupling larger than the value required by the unitarity condition, in order to explain the diphoton excess.  
Note that if we lose the perturbative condition as $\lambda_{H \phi^Q} \lesssim 4 \pi$, then the cases of $N=5$ and $N=7$ also have allowed parameter space.
However, we need careful analysis to find a parameter space that satisfies perturbative unitarity,
which is beyond the scope of this work.

%%%%%%%%%%%%%%%%%%%%%%%%%%%%%%%%%%%%%%%%%%%%%%%%%%%%%%%%%%%%%%%%%%%%%%%%%%%%%%%%%%%%%%%%%%%%%%%%%%%%%%%%%

%\section{Conclusions}
\section{ Conclusions and discussions}
We have proposed a new three-loop induced radiative neutrino model with local dark $U(1)$ symmetry, 
in which the discrepancy of the muon anomalous magnetic moment within the standard model can be 
resolved by using exotic charged fermions, and both DM candidates (the Majorana fermion and/or scalar) 
%{\color{red} pseudoboson (??) WHAT IS PSEUDOBOSON ???}) 
can satisfy the observed thermal relic density   without conflict with the results of direct detection searches, 
considering that the DM pair annihilation into a pair of 
$Z'$ bosons is supposed to be the dominant process $XX\rightarrow Z' Z'$. 
% to explain the relic density.
%%%
We have also generalized the hypercharges of isospin doublet fields as well as isospin singlet fields without 
violating the structure of neutrino masses at the three-loop level. 
As a result, a lot of electrically charged new fields can be involved in our theory. 
In this case, such a general value of hypercharge could cause a stability problem; therefore, we have to make 
them decay into the SM (or DM) appropriately. In order to realize this, we have added some more nonzero 
charged bosons, and have shown the appropriate decay processes for each value of hypercharge, retaining our 
model structure for the neutrino masses and  mixings. 
Here such new bosons also play a role in contributing the diphoton excess  
at 750 GeV that was reported  recently by both ATLAS and CMS collaborations.
%%%

Then, we have investigated the production of the 750 GeV scalar particle $H$ which appears as a linear combination 
of the SM Higgs and a neutral CP even component of the $U(1)$ charged SM singlet scalar.
This scalar particle $H$ is produced by gluon fusion via mixing with SM Higgs, and also by the photon fusion process.
We find that a $3-10$ fb cross section for $pp \to H \to \gamma \gamma$ can be obtained by $O(1)$ TeV trilinear coupling for $H$ and charged scalar,   which is safe from tree level unitarity.
The decay width of $H$ is $O(10)$ GeV due to the contribution from the $H \to Z' Z'$ mode, where a larger gauge coupling would generate a larger width of $H$.
Moreover, we have shown that the constraint from a diphoton search at 8 TeV can be satisfied.
Thus, we have explained the diphoton excess naturally, depending on the number of 
hypercharge for new isospin doublet scalar field. 

Before closing, we would like to emphasize that the three-loop  radiative neutrino mass 
model presented in this paper is new and has its own value even if the 750 GeV diphoton 
excess goes away in the future.    The three-loop diagrams relevant for the neutrino masses 
within this model are topologically different from the previous models in the literature, if we 
trace the dark charge flows in the Feynman diagrams.  The model would remain as an 
interesting and viable model for radiative neutrino masses and also for the muon $(g-2)_\mu$.
As such, it deserves its own investigation at current and future colliders, and in low energy lepton flavor physics.

%\section*{ Appendix}
%%%%%%%%%%%%%%%%%%%...

%\newpage
%%%%%%%%%%%%%%%%%%%%%%%%%%%%%%%%%%%
%\hspace{0.2cm} {\bf Acknowledgments}
%\section*{Acknowledgments}:
%\vspace{0.5cm}
\section*{Acknowledgments}
\vspace{0.5cm}
H.O. thanks Shinya Kanemura, Kenji Nishiwaki,  Seong Chan Park,  Ryoutaro Watanabe 
and Kei Yagyu for fruitful discussions. 
This work is supported in part by National Research Foundation of Korea (NRF) Research 
Grant No. NRF-2015R1A2A1A05001869 (PK), and by the SRC program of NRF Grant No. 20120001176 funded by MEST through 
the Korea Neutrino Research Center at Seoul National University (P.K., Y.O.).
%H. O. is sincerely grateful for all the KIAS members, Korean cordial persons, foods, culture, %weather, and all the other things.
%%%%%%%%%%%%%%%%%%%%%%%%%%%%%%%%%%%
%%%%%%%%%%%%%%%%%%%%%%%%%%%%%%%%%%%

\section*{Appendix}
%%%
Here we explicitly show the loop functions $G_{I-IV}$ that appear in the neutrino sector:
\begin{align}
&G_I (x_I)=%\sum_{a=R,L}
\int\Pi_{i=1}^3 dx_i \frac{\delta(\sum_{i=1}^3 x_i-1)}{(x_3^2-x_3)^2}
\int\Pi_{i=1}^4 dx'_i \frac{\delta(\sum_{i=1}^4 x'_i-1)}{({x'}_4^2-x'_4)^2}
\int\Pi_{i=1}^3 dx''_i {\delta(\sum_{i=1}^3 x''_i-1)}\times
\nn\\&
\frac{x_1''}
{\left[
x''_2 X_{E^{-5}_\alpha} + x''_3 X_{S^{-5}} -\frac{x''_1}{{x''}^2_4 -x''_4}
\left(
x'_2 X_{\Psi_{I\alpha}} + x'_3 X_{S_R}  + x'_4 X_{\phi^{-4}}  -\frac{x'_1}{x_3^2-x_3}
(x_1 X_{E^{-5}_\gamma} +x_2 X_{S^{-5}} + x_3 X_{\phi^{-4}} )
\right)
\right]^2},\nn\\
&-(S_R\to S_I),\\
%%%
&G_{II} (x_I)=\int\frac{\Pi_{i=1}^4 dx_i \delta(\sum_{i=1}^4 x_i-1)}{(x_3 + x_4)^2 (x_3 + x_4 -1)^2}
\int \frac{\Pi_{i=1}^3 dx'_i \delta(\sum_{i=1}^3 x'_i-1)}{[ (a{x'}_1+ x'_3)^2 -a^2 x'_1-x'_3]^2} \times
%\int\Pi_{i=1}^3 dx''_i {\delta(\sum_{i=1}^3 x''_i-1)}
\nn\\&
\int \frac{\Pi_{i=1}^3 dx''_i {\delta(\sum_{i=1}^3 x''_i-1)} x''_1 }
{\left[x''_2 X_{E^{-5}_\alpha} + x''_3 X_{S^{-5}} - \frac{x''_1(x'_2 X_{\Psi_{I\beta}} +x'_3 X_{\phi^{-5}} -x'_1 c)}{(a x'_1 +x'_3)^2 - a^2 x'_1-x'_3}\right]^2\left(1-\frac{x_1'x_1'' b}
{(a x'_1 +x'_3)^2 - a^2 x'_1-x'_3}\right)^2}-(S_R\to S_I),\\
%%%
&G_{III} (x_I)=\int\frac{\Pi_{i=1}^4 dx_i \delta(\sum_{i=1}^4 x_i-1) x_2^3}{(x_1^2 - x_1)^2 (x_1 -1)^3}
\int \frac{\Pi_{i=1}^3 dx'_i \delta(\sum_{i=1}^3 x'_i-1)}{ ({x'}_1-1 )^2} \times
%\int\Pi_{i=1}^3 dx''_i {\delta(\sum_{i=1}^3 x''_i-1)}
\nn\\&
\int \frac{\Pi_{i=1}^3 dx''_i {\delta(\sum_{i=1}^3 x''_i-1)} x''_1 }
{(x''_2 X_{E^{-5}_\alpha} + x''_3 X_{S^{-5}} - x''_1 d) \left(1- \frac{x''_1}{1-x'_1}\right)^3}
-(S_R\to S_I),\\
%%%
&G_{IV} (x_I)=\int\frac{\Pi_{i=1}^4 dx_i \delta(\sum_{i=1}^4 x_i-1)}{(x_2 + x_4)^2 (x_2 + x_4 -1)^2}
\int \frac{\Pi_{i=1}^3 dx'_i \delta(\sum_{i=1}^3 x'_i-1) {x'}^2_1 B}{ D^2} \times
%\int\Pi_{i=1}^3 dx''_i {\delta(\sum_{i=1}^3 x''_i-1)}
\nn\\&
\int \frac{\Pi_{i=1}^3 dx''_i {\delta(\sum_{i=1}^3 x''_i-1)} x''_1 }
{x''_2 X_{E^{-5}_\alpha} + x''_3 X_{S^{-5}} - x''_1 F }
-(S_R\to S_I),
%%%
\label{mnu1}
\end{align}%\end{widetext}
%%%
with
\begin{align}
a&\equiv \frac{x_4(x_3+x_4+1)}{(x_3 + x_4) (x_3 + x_4 -1)} ,\quad
b\equiv a^2-\frac{x^2_4 -x_4}{(x_3 + x_4) (x_3 + x_4 -1)},\\
c&\equiv \frac{x_1 X_{E^{-5}_\gamma} +x_2 X_{S^{-5}} +x_3 X_{\phi^{-5}} +x_4 X_{S_R}}{(x_3 + x_4) (x_3 + x_4 -1)},\\
%%%
d&\equiv \frac{(1-x_1)^2 (x'_2 X_{E^{-5}_\gamma} +x'_3 X_{S^{-5}}) }{x_2^2 ({x'}^2_1 -x'_1)}
-\frac{(x_1 -1)\left[ (x_1+x_2)X_{\phi^{-5}} +x_3 X_{\Psi_{I\beta}} +x_4 X_{S_R}\right] }{x_1 x_2^2 (x'_1 -1)},\\
%%%
B& =\frac{(x_2 + x_4)(x_3 + x_4)-x_4}{(x_2 + x_4)(x_2 + x_4-1)},\\
D&=({x'}^2_1 -x'_1) B^2 +x'_1
\left[
\left( \frac{(x_2 + x_4)(x_3 + x_4)-x_4}{x_2 + x_4-1} \right)^2 - \frac{(x_3 + x_4)(x_3 + x_4-1)}{(x_2 + x_4)(x_2 + x_4-1)} \right],\\
F&=
\frac1D
\left[
x'_2 X_{E^{-5}_\gamma} +x'_3 X_{S^{-5}} -\frac{x'_1 [x_1 X_{\Psi_{I\beta}} + (x_2+x_3) X_{\phi^{-5}}+x_4 X_{S_R}]}{(x_2 + x_4)(x_2 + x_4-1)}
\right],
\end{align}
where $m_{R}$ and $m_{I}$ are the masses of $S_R$ and $S_I$ and satisfy $m_R^2-m^2_I=\mu v'/(2\sqrt2)$ 
and we define  $X_f\equiv (m_f/M_{\rm max})^2$, and $M_{\rm max}={\rm Max}[M_L,M_{\Psi_i}, m_{S^\pm},m_{S^{\pm5}},m_{R},m_{I}]$.
%%%

\end{document}